\documentclass{egpubl}
 
\JournalSubmission

\usepackage[T1]{fontenc}
\usepackage{dfadobe}  

\usepackage{cite}  \BibtexOrBiblatex
\electronicVersion
\PrintedOrElectronic
\ifpdf \usepackage[pdftex]{graphicx} \pdfcompresslevel=9
\else \usepackage[dvips]{graphicx} \fi

\usepackage{egweblnk}

\title[A Dataset and Benchmark for Mesh Parameterization]{A Dataset and Benchmark for Mesh Parameterization}

\author[G. Shay et al.]
{\parbox{\textwidth}{\centering G. Shay$^{1}$,
    J. Solomon$^{1}$,
    and O. Stein$^{1}$
        }
        \\
{\parbox{\textwidth}{\centering $^1$Massachusetts Institute of Technology, MA, USA
       }
}
}

\usepackage{xcolor}
\usepackage{amsmath, amsfonts}
\newcommand{\R}[0]{\mathbb R}

\begin{document}

\teaser{
 \includegraphics[width=\linewidth]{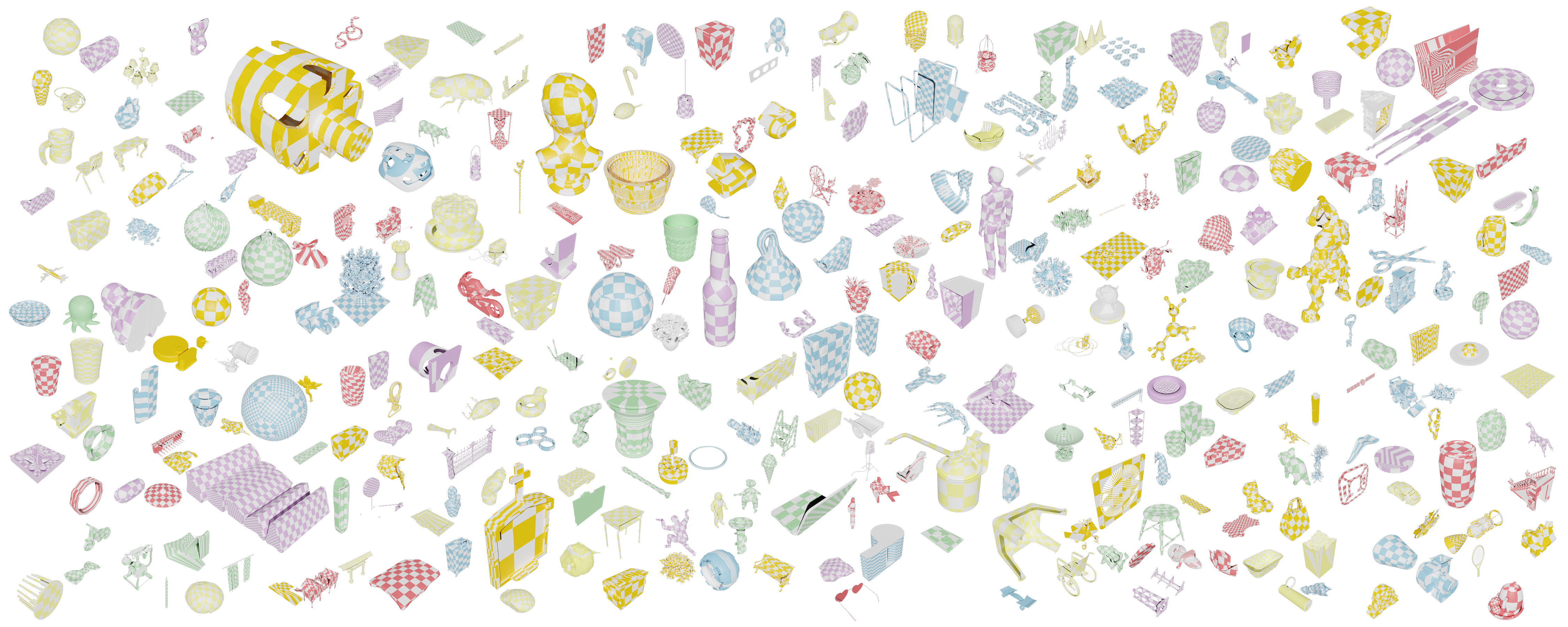}
 \centering
  \caption{A tiled representation of our dataset.
  The meshes
  and their UV maps were created by digital artists.
  They are representative of the the challenges that parameterization algorithms
  face in practice. Using our benchmark, the
  artist-provided UV maps can be directly compared to the UV maps computed by an automatic parameterization method.}
\label{fig:teaser}
}

\maketitle
\begin{abstract}
    UV parameterization is a core task in computer graphics, with applications in mesh texturing, remeshing, mesh repair, mesh editing, and more.
    It is thus an active area of research, which has led to a wide variety of parameterization methods that excel according to different measures of quality.
    There is no single metric capturing parameterization quality in practice, since the quality of a parameterization heavily depends on its application; hence, parameterization methods can best be judged by the actual users of the computed result.
    In this paper, we present a dataset of meshes together with UV maps collected from various sources and intended for real-life use.
    Our dataset can be used to test parameterization methods in realistic environments.
    We also introduce a benchmark to compare parameterization methods with artist-provided UV parameterizations using a variety of metrics.
    This strategy enables us to evaluate the performance of a parameterization method by computing the quality indicators that are valued by the designers of a mesh.

\ccsdesc[500]{Computing methodologies~Texturing}

\printccsdesc   
\end{abstract}

\section{Introduction}

\emph{Mesh parameterization} is a core task in geometry processing.
A parameterization (also called a \emph{UV map}) is a map from a triangle mesh in
$\R^3$ to the plane $\R^2$.
UV maps produced with parameterization algorithms are used to store
high-frequency detail
such as images, normal maps, and bump maps, that exist in 2D parametric domains.
Nearly all standard software tools for 3D modeling and design include
automatic tools for parameterization.

Given their popularity and broad applicability, a variety of automatic
algorithms for parameterization
exist.
These algorithms have very different goals and navigate different trade-offs.
Some aim to minimize a diverse variety of distortion measures, some aim
to pack a parameterization tightly into a bounded box, and some aim for
application-specific goals.
Any subset of these considerations might be important for a given application,
and hence there is no single agreed-on criterion with which to judge
parameterizations.
As a result, parameterization methods are evaluated inconsistently in the research
literature.
Different datasets are used, often containing surfaces that were not intended
to be parameterized in the first place
(but instead, 3D printed \cite{Thingi10K}),
and quantitative evaluations use different metrics across articles.
This situation makes it difficult to compare parameterization methods to each
other, to identify challenging cases for future work, and to evaluate
methods for practical application.

In an effort to address this situation,
we propose a dataset and benchmark for evaluating mesh parameterization
algorithms.
Our goal is not to present a one-dimensional leaderboard to rank algorithms,
but rather to make it easy to understand (1) the distinctive properties of a
parameterization technique and (2) how that technique differs from its peers and
from hand-made parameterizations generated by artists.

Our dataset is constructed from
337 3D models gathered from online repositories for digital art, such as
Blenderkit \cite{blenderkit} and Sketchfab \cite{sketch-fab}.
The dataset is publicly available, and we include proper licensing information
for each object.
This dataset is
constructed
to be as realistic as possible:
Most of the models are sourced from online artistic and 3D printing repositories
rather than from geometry processing research papers, and they are all
accompanied by artist-designed parameterizations.
Testing parameterization algorithms on this dataset
characterizes the behavior of a method in a realistic application, especially in
edge cases that might not be included in synthetic datasets but occur in
practice.

Our benchmark evaluates the performance of a parameterization method applied
to our dataset.
The benchmark produces plots of major quantities of interest that characterize
the distortion, the resilience, and the practical usability of the produced
UV maps.
We include both geometric measurements conventionally used in research papers
to evaluate parameterization algorithms, as well as measurements demonstrating
how parameterization algorithms compare to artist-created UV maps.

The
benchmark code
is included with this publication and can be readily applied to
future parameterization methods.

\section{Related Work}

\subsection{Mesh Datasets}

Datasets of surfaces have been created to evaluate many tasks in geometry processing.
For example, SHREC \cite{shrec} is a series of datasets with a relatively small number of objects used to test object retrieval methods.
Similarly, the Princeton Shape Benchmark \cite{princeton-shape-benchmark} contains 1,814 models intended for object retrieval.
TOSCA \cite{tosca} contains 80 surfaces in a variety of poses, constructed to test shape correspondence methods.
The FAUST dataset \cite{faust} contains 300 scans of real humans.
Vlasic et al.~\cite{mitanimation} provide a dataset of animated humans with correspondence.

Thingi10k \cite{Thingi10K} contains ten thousand meshes created
by users of Thingiverse mostly for the purpose of 3D printing.
Like our dataset, Thingi10k can be used to test geometry processing algorithms
with a large number of real-world targets used by actual end users.
Although it was not constructed to test parameterization methods specifically,
Thingi10k was used to test at least one parameterization algorithm
\cite{discrete-conformal}.

ABC \cite{abc-dataset} is a large dataset of CAD models intended for a variety
of applications in geometric deep learning.
Other learning-oriented datasets include ShapeNet \cite{shapenet},
ModelNet \cite{modelnet}, and PartNet \cite{partnet}.
Some machine learning papers introduce their own datasets, e.g.,\ \cite{meshcnn}.

A few past datasets were collected specifically to test parameterization
algorithms. 
Myles et al.'s dataset \cite{robust-field-aligned} of 114 meshes---which builds
on AIM@Shape \cite{AimAtShape} and the Stanford model repository
\cite{stanford-repository}---is built to test mesh parameterization algorithms.
Similarly, Liu et al.~\cite{progressive-parameterizations} introduce a dataset
of 20,712 disk topology meshes for parameterization algorithms, with cuts
automatically inserted;
they also randomly distort meshes to make parameterization more challenging.
Liu et al.'s dataset has been used to test other recent parameterization methods
\cite{wrapd,flip-free-param}.
While these datasets can be used to expose geometric properties and
efficiency of assorted parameterization algorithms, they do not consist
exclusively of real-world objects meant to be textured, they do not contain
hand-created UV maps for comparison, and they do not feature meshes of widely
varying topology.
Hence, these past datasets are not strong proxies for real-world data,
reflecting the demands on parameterization software deployed in practice.

\subsection{Benchmarks for Geometry Processing Tasks}

Chen et al.~\cite{seg-bench} introduce a dataset and benchmark for mesh segmentation.
Like our work, they include artist-created ground truth segmentations that can be used to evaluate results of an automatic mesh segmentation algorithm.
Similarly to our setting, their benchmark does not consist of a single metric along which a method is judged, but rather automatically compares to the manually-designed segmentations on a variety of axes.

Wang et al.~\cite{collision-benchmark} present a benchmark for continuous collision detection that compares collision algorithms to the ground truth, derived using symbolic calculation.
Nehm\'{e} et al.~\cite{texturedmeshqualityassessment} introduce a dataset and benchmark for texture quality assessment.
They use a perceptual loss based on features from a pre-trained deep image-based network.
The dataset is constructed from 55 meshes from SketchFab.

Many geometry processing research articles contain ad hoc benchmarking strategies to assess their method.
For parameterization, typical benchmark measurements include the number of flipped triangles, assorted measures of distortion (often the one their method directly optimizes), and runtime.
Parameterization quality is also often judged by eye with the display of checkerboard patterns on the surface, similar to Fig.~\ref{fig:teaser} (see \S\ref{sec:parameterizationmethods} for references).

\subsection{Parameterization Methods}
\label{sec:parameterizationmethods}

The task of constructing UV maps goes back to the earliest days of computer
graphics \cite{catmullphdthesis} and is still an active area of research.
Given the importance of UV parameterization in geometry processing, it comes as
no surprise that a variety of algorithms have been proposed for this task. 

Most recent research papers in parameterization focus on a single piece of the parameterization pipeline---e.g., cutting a mesh or flattening a patch with disk topology---rather than considering the entire problem
of placing seams, cutting, and flattening.
A particularly attractive problem has been that of finding a \emph{flip-free} (or, injective) map of a mesh into the plane, after it has been cut into pieces of disk topology.  In this problem, one seeks a parameterization in which no triangle has been inverted and/or a map where no two points of the surface are mapped to the same coordinate in the UV plane (global injectivity). Several surveys review past and current parameterization methods, including \cite{tutorial-survey,survey-planar,inversion-free-survey}.

In \S\ref{sec:casestudy}, we  evaluate a few current methods for
parameterization as examples of the diversity of approaches in this field.
Least-squares conformal mapping \cite{lscm} (LSCM) is a classical parameterization method that finds a UV map by solving a linear system of equations measuring conformal energy.
LSCM is efficient but does not cut the mesh and does not guarantee injectivity. 
Rabinovich et al.~\cite{slim} (SLIM) compute a flip-free map by optimizing the symmetric Dirichlet distortion energy \cite{Smith2015,Schreiner2004} starting from an injective parameterization.
They do not cut the mesh, and the provided implementation works only on disk topology meshes.
Garanzha et al.~\cite{foldover-free} provide an alternative to SLIM for computing flip-free maps.
Their implementation requires a choice of initial UV map for the algorithm; we initialize with LSCM for our case study (as suggested by the authors). Their method also does not cut meshes.
Gillespie et al.~\cite{discrete-conformal} compute a flip-free conformal map, but they remesh the surface in the process---without changing the geometry.
Finally, unlike the other methods in our case study, Optcuts \cite{optcuts} cuts a mesh to the correct topology before flattening it.
Our dataset contains test cases specifically for methods like these, and the benchmark measures cut length to evaluate different cutting strategies.

\begin{figure}
    \centering
    \includegraphics[height=0.4\linewidth]{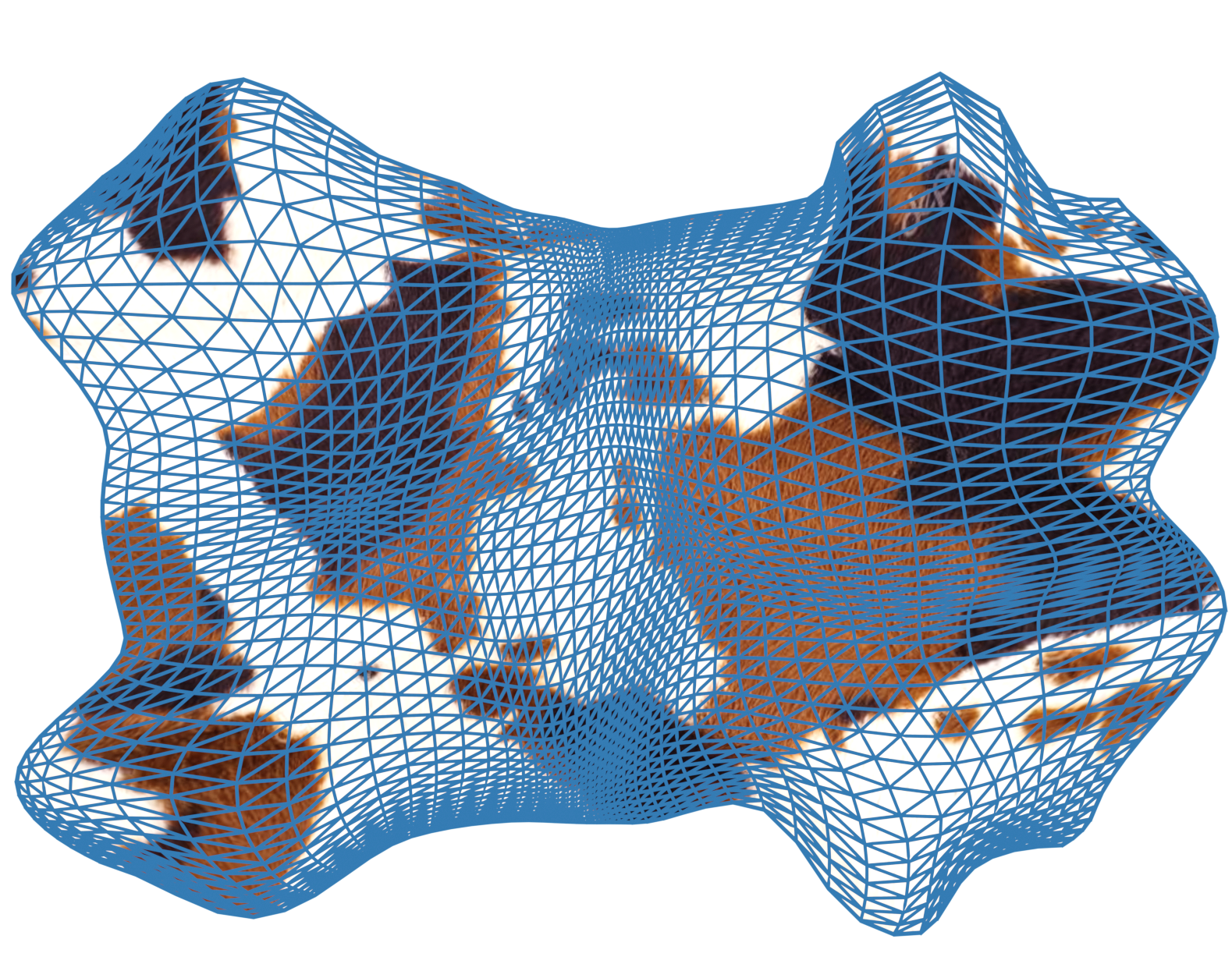}
    \includegraphics[height=0.42\linewidth]{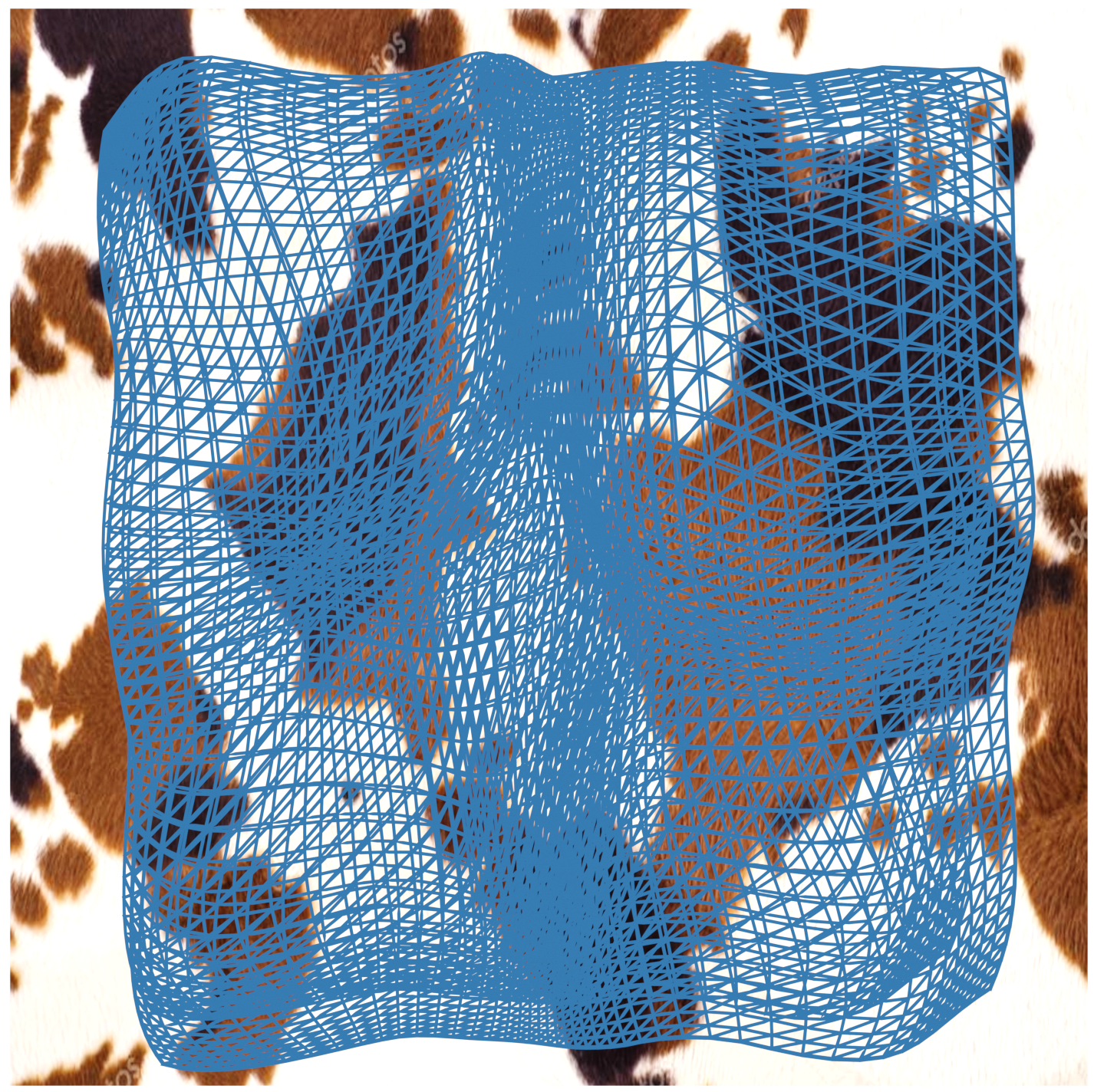}
    \caption{A textured cowhide mesh with triangles on both sides and thus disk topology (\emph{left}).
    The artist chose a UV map (\emph{right}, resized to fit in unit square) that is not globally injective, as this was apparently not important for their application (both sides have a very similar texture).
    The artist also decided to distort the UV map, even though a distortion-free map would have been possible.}
    \label{fig:cowhide}
\end{figure}

\section{Designing a Benchmark}

With the wide diversity of parameterization methods and their applications
in mind, we
seek
to create a benchmark able to accurately identify important features of a method's UV map and compare to the desires of digital artists.

The following desiderata inform the design of our dataset and benchmark:
\begin{enumerate}
    \item We want to represent realistic, real-world applications in digital art.
    \item Our benchmark should contain simple and general metrics capturing the degree to which the texture map is distorted when laid on the mesh.
    \item The dataset and benchmark should account for the diversity of parameterization algorithms and their different goals.
\end{enumerate}

The meshes used to test parameterization methods are often not the kinds of meshes that users actually want to texture---researchers take surfaces, place arbitrary seams on them, and then flatten the surface without actually planning to texture them.
In comparison, we aim to compare how well automatic parameterization methods correspond to parameterizations designed by artists.
We achieve this by building our dataset from real-world meshes that were UV mapped and textured by artists
(\S\ref{sec:sources}) and by comparing a method's UV maps to the artists' designs.
Since our dataset contains real-world meshes, some of our examples do not fit the assumptions made by parameterization algorithms; for instance, they can be nonmanifold, can have complicated topologies, and can have triangulations ill-suited for numerical computation.

Our dataset includes artist-designed texture maps, which are valuable for benchmarking parameterization methods.
Our accompanying benchmarking method includes metrics carefully-designed to uncover the intent of the artist when compared to algorithmically-design parameterizations.
For instance, our benchmark includes a correlation of per-triangle distortion to the artist.
This accounts for the fact that creators may decide to concentrate distortion on non-visible or semantically unimportant parts of a model.
Artists might also decide to only care about injectivity for certain parts of the model.
Fig.~\ref{fig:cowhide} shows an example of a rug that is textured on both sides with a similar texture.
Instead of putting both sides of the rug next to each other (assuring global injectivity), the author chose to overlap both sides in UV space, perhaps since they are textured with the same image anyways.
Instead of choosing a zero-distortion map, which would be possible due to the planarity of the mesh, we speculate that the author chose to distort the mesh to better pack it into a rectangle.

More broadly, 
the quality of UV maps is notoriously hard to quantify.
As a result, UV mapping articles have used a wide variety of distortion metrics to judge the quality of their parameterizations.
Because of this, we
compare the parameterizations on a variety of simple, standard measures that an artist could eyeball, like 
flipped triangles, area distortion, and angle distortion,
as well as more complicated measures, like the
symmetric Dirichlet energy (\S\ref{sec:measures}).

Not all UV parameterization methods have the same goal.
Some aim at producing a flip-free map, some aim at reducing distortion. Some only work on disk-topology surfaces, some create their own seams. Some might aim for non-manifold meshes, while others operate on manifold meshes.
Our dataset contains a wide array of meshes of all types
(\S\ref{sec:datasetproperties}), and our benchmark contains many different measures to account for this diversity (\S\ref{sec:measures}).

\section{Dataset}

\begin{figure}
    \centering
    \includegraphics[width=\linewidth]{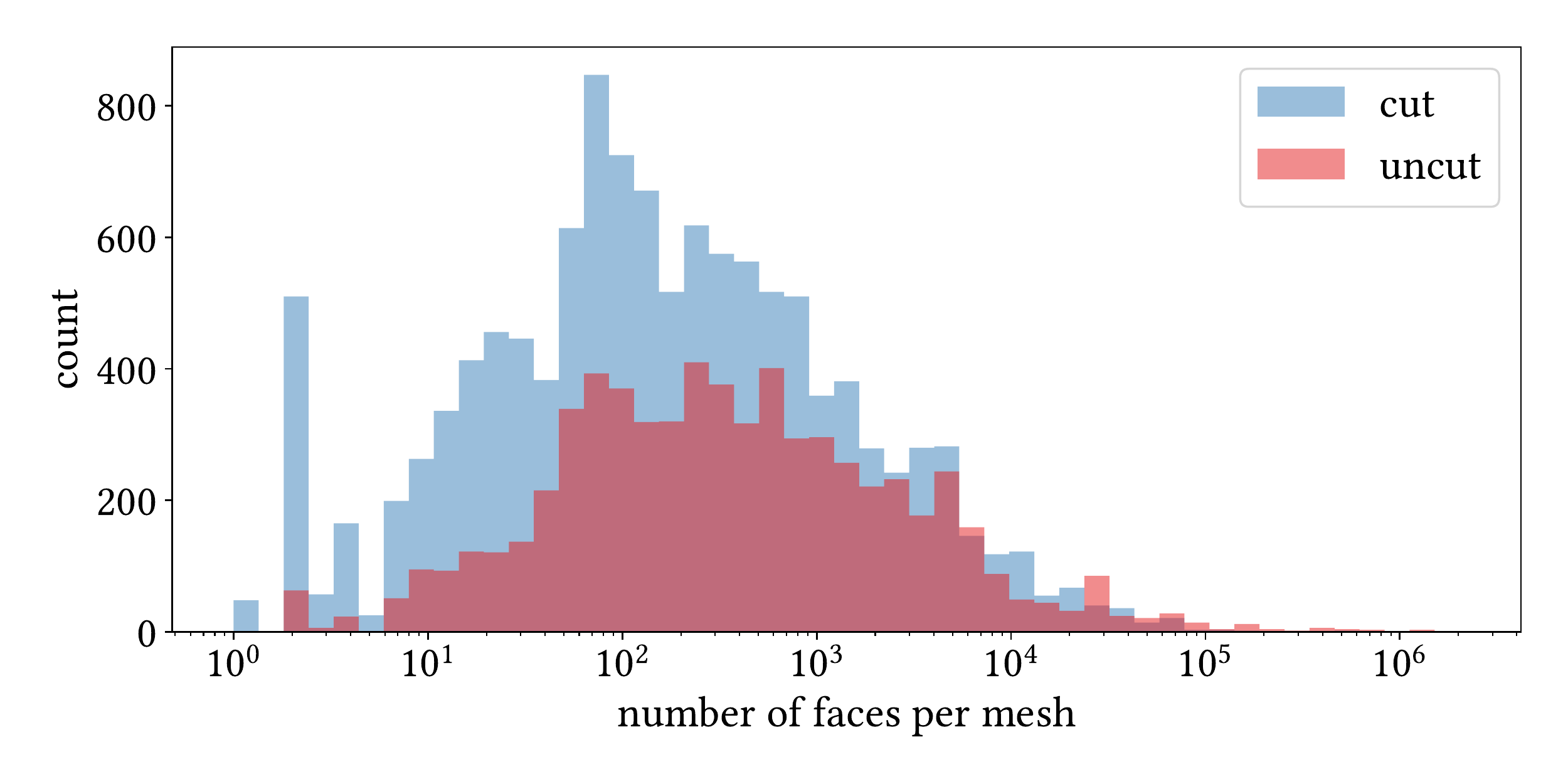}
    \caption{
    The distribution of number of faces among the uncut and the cut portions of our dataset.
    }
    \label{fig:faces}
\end{figure}

We begin our technical discussion by detailing how we gathered our dataset of meshes with hand-designed parameterizations.  We also report summary statistics describing its properties.

\subsection{Sources}
\label{sec:sources}

We sourced our meshes from Blenderkit ~\cite{blenderkit}, Sketchfab ~\cite{sketch-fab}, Thingiverse ~\cite{thingiverse}, PolyHaven \cite{poly-haven}, and Keenan Crane's model repository \cite{keenan-crane}. The meshes on these websites are used for a variety of purposes, including for animation, for 3D printing, and as video game assets. The included texture maps are thus representative of the types of parameterization used in many different domains. We have reviewed the copyright for each mesh before inclusion in the dataset and provide the license information for each mesh in
the supplemental material.

\subsection{Preprocessing}

Each mesh is run through a preprocessing step in Blender \cite{blender} to make it suitable for automatic parameterization algorithms.
This step includes
\begin{enumerate}
    \item triangulation,
    \item merging of close vertices (close in 3D \emph{and} UV space), and 
    \item splitting into connected components and discarding all but the 50 largest components.
\end{enumerate}
This procedure forms the \emph{uncut} version of the dataset: a parameterization method would need to create its own seam to compute a UV map.

Since not every parameterization algorithm includes a strategy for cutting meshes into patches of proper topology, we also create a \emph{cut} version of the dataset by applying the artist's seams as cuts along the mesh.
In the cut dataset, we also exclude nonmanifold meshes, remove unreferenced vertices, and remove meshes where the connectivity of the surface does not correspond to the connectivity of the UV map.
Again only the 50 largest components per asset are included.

\begin{figure}
    \centering
    \includegraphics[width=0.3\linewidth]{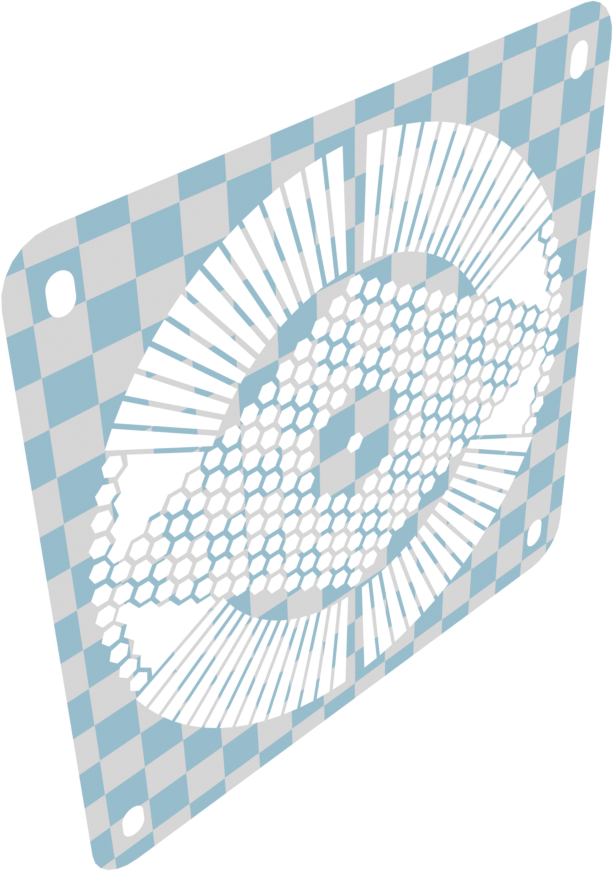}
    \includegraphics[width=0.65\linewidth]{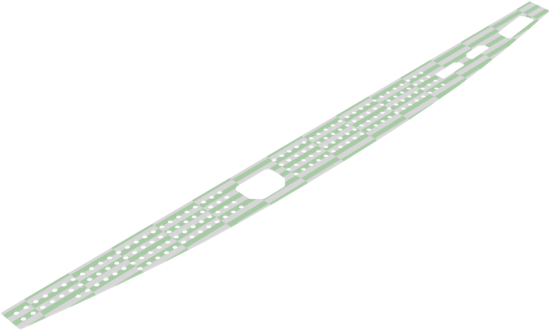}
    \caption{
    A mesh from our dataset with genus 291 \emph{(left)}, and
    a mesh with 357 boundary segments \emph{(right)}.
    }
    \label{fig:outlier}
\end{figure}

\subsection{Tags}
\label{sec:dataset-tags}

In addition to creating the \emph{cut} and \emph{uncut} versions of our dataset,
we also tag the meshes with a variety of labels.
Different methods operate on different kinds of shapes---some methods only
compute parameterizations for disk topology meshes, while other methods can
handle more complicated topologies.
There are methods that can segment meshes that are not flattenable in order to make
parameterization possible, while other methods can even segment non-manifold
meshes.
Our labels are:
\begin{itemize}
    \item \emph{Disk:} whether or not the mesh has Euler characteristic 1
    \item \emph{Closed:} whether or not the mesh is closed (i.e., has no boundary)
    \item \emph{Manifold:} whether or not the mesh is manifold: both vertex-manifold (the faces adjacent to a vertex form a fan) and edge-manifold (no edge is adjacent to more than two faces). All meshes in the cut dataset are manifold.
    
    \item \emph{Small:} whether or not the mesh has fewer than 100 faces
\end{itemize}

\noindent The benchmark uses switches on the command line to allow a user to run it on
 any subset of these four tags.

\begin{figure}
    \centering
    \includegraphics[width=\linewidth]{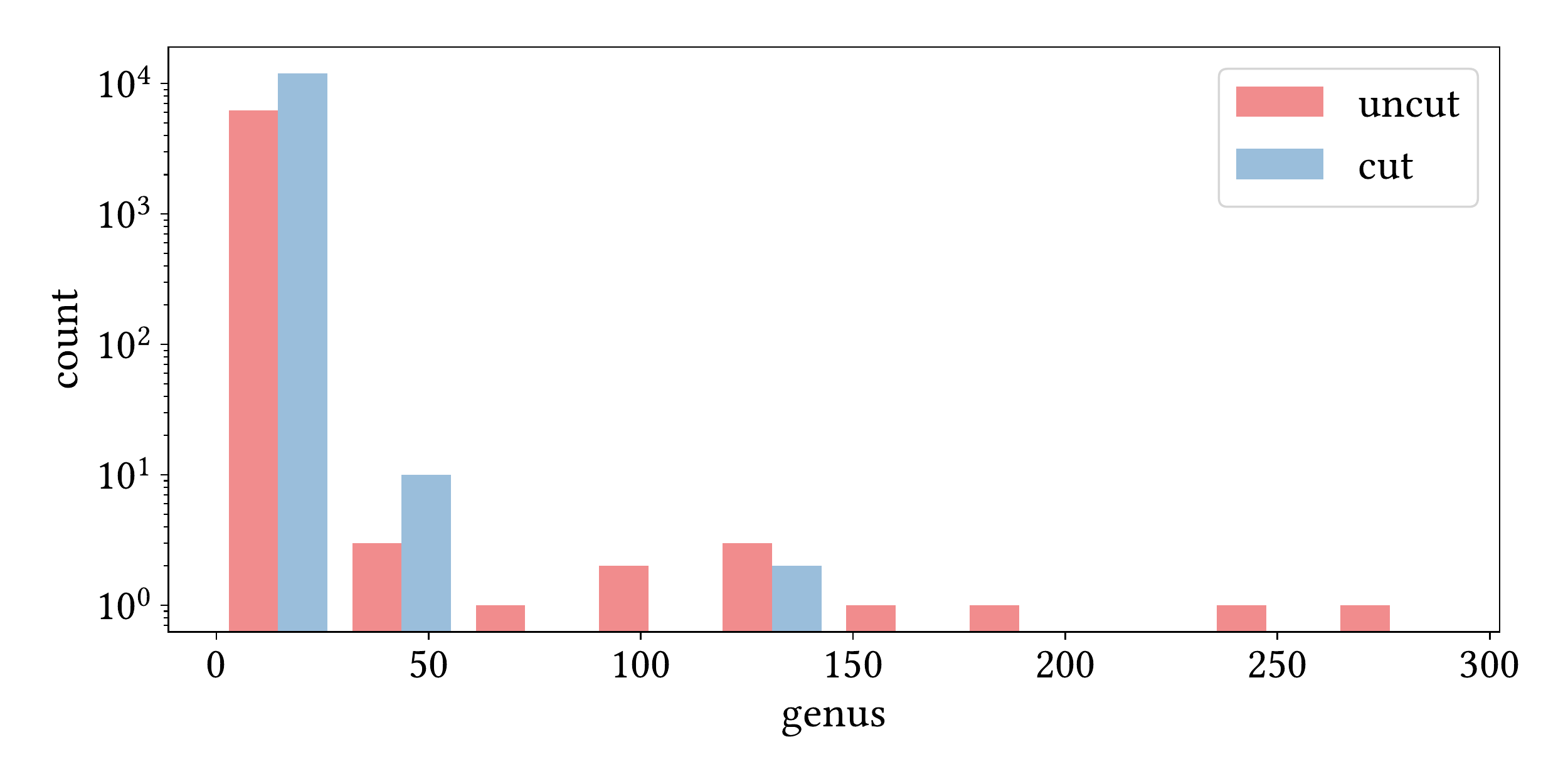}
    \includegraphics[width=\linewidth]{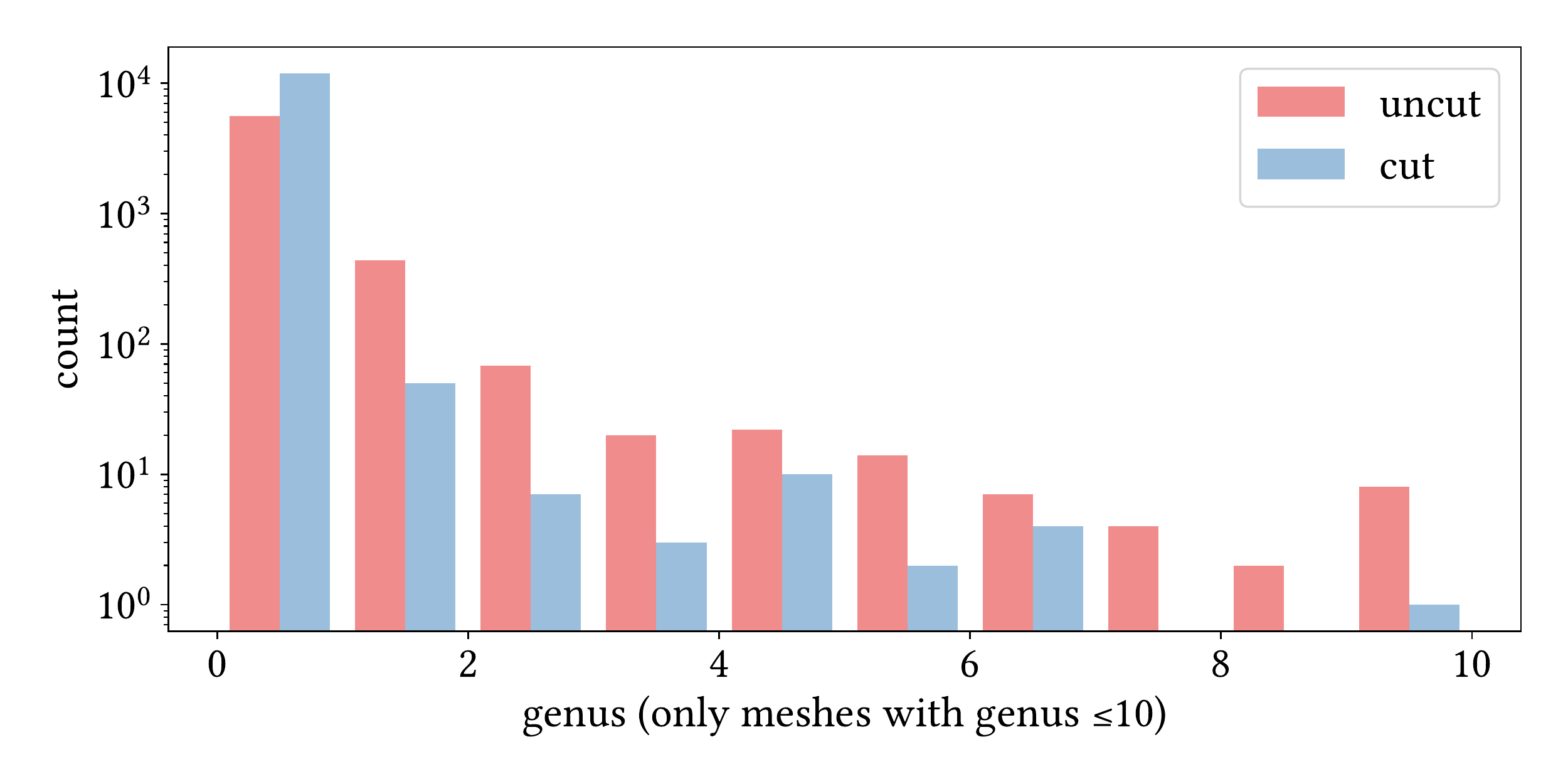}
    \caption{
    Genus of meshes in the entire dataset \emph{(top)} and of meshes
    with genus less than or equal to 10 \emph{(bottom)}.
    }
    \label{fig:genus}
\end{figure}

\subsection{Statistics}
\label{sec:datasetproperties}

\begin{figure}
    \centering
    \includegraphics[height=0.5\linewidth]{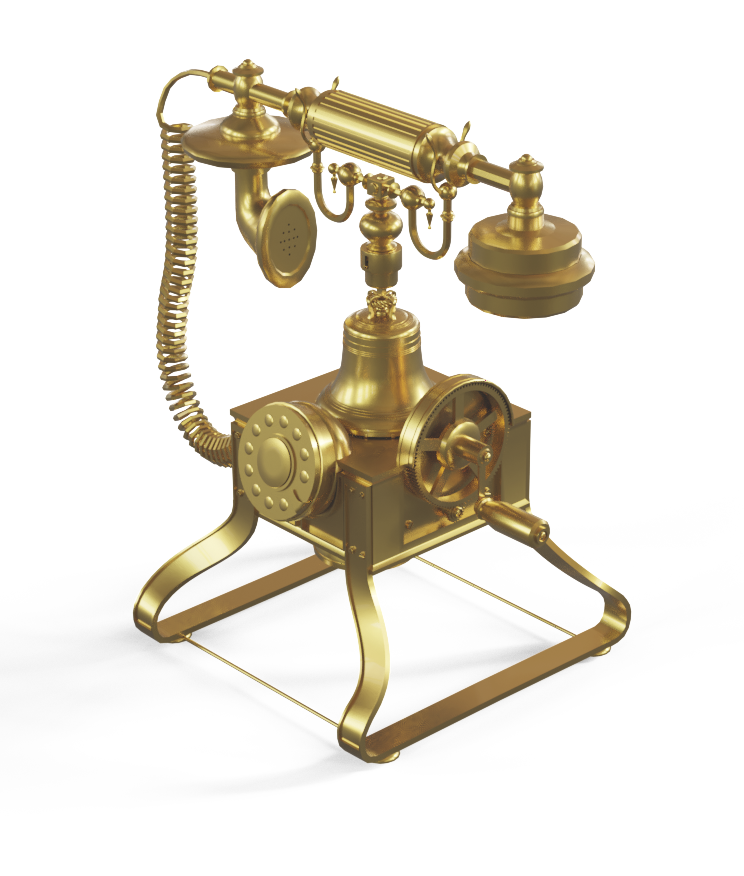}
    \includegraphics[height=0.5\linewidth]{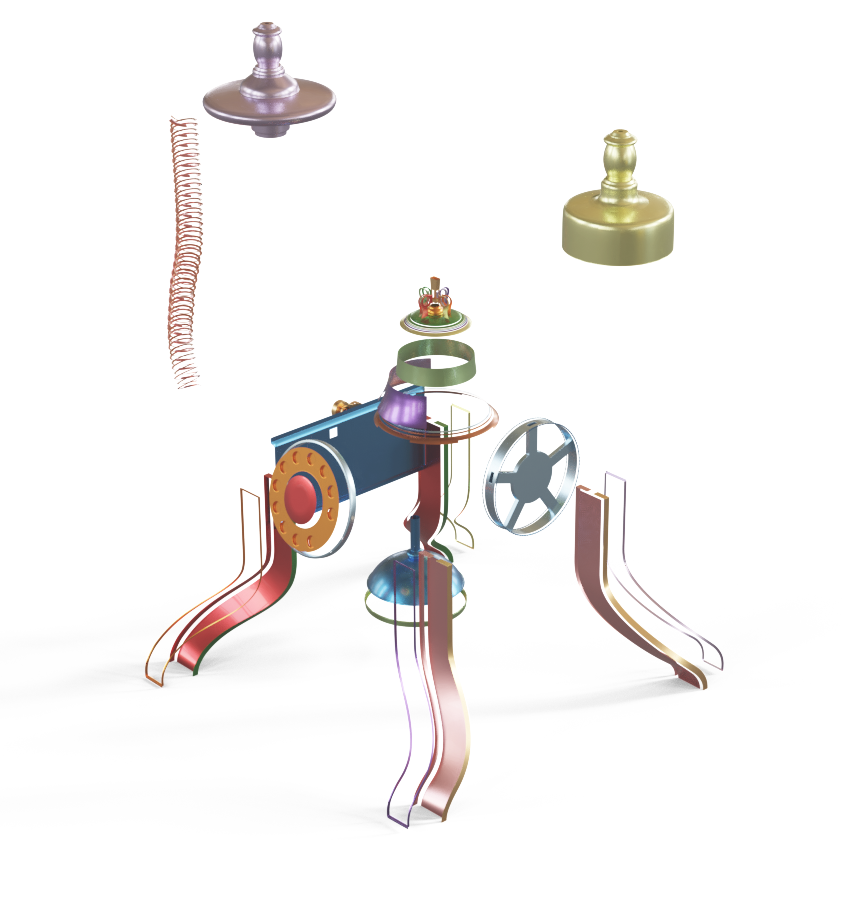}
    \caption{
    One of the 337 artist-generated assets that served as the basis
    for our dataset (\emph{left}), and some of the cleaned connected components
    that are meshes in our dataset (\emph{right}).
    }
    \label{fig:allparts}
\end{figure}

We here list a few statistics about our dataset.
Our dataset consists of 11,913 cut meshes and 6,478 uncut meshes.
These meshes are generated from 337 manually-selected source meshes. 
Fig.~\ref{fig:allparts} shows how one complicated artist-generated object results
in many cleaned connected meshes.
The median number of faces per mesh is 192, with 142 for the cut meshes and 312 for the uncut meshes; see Fig.~\ref{fig:faces} for the distribution of mesh sizes in the dataset.

The median number of charts in the dataset per source mesh is 26.5, with 50 for the cut meshes and 8 for the uncut meshes. We capped the number of charts outputted per source mesh at 50 to enforce a greater variety within our dataset.

Most meshes in the dataset have a genus smaller or equal to 10, with a few
outliers (see Fig.~\ref{fig:genus}).
Fig.~\ref{fig:outlier} shows an example of such an outlier with an
unusually large genus of 291.

\begin{figure}
    \centering
    \includegraphics[width=\linewidth]{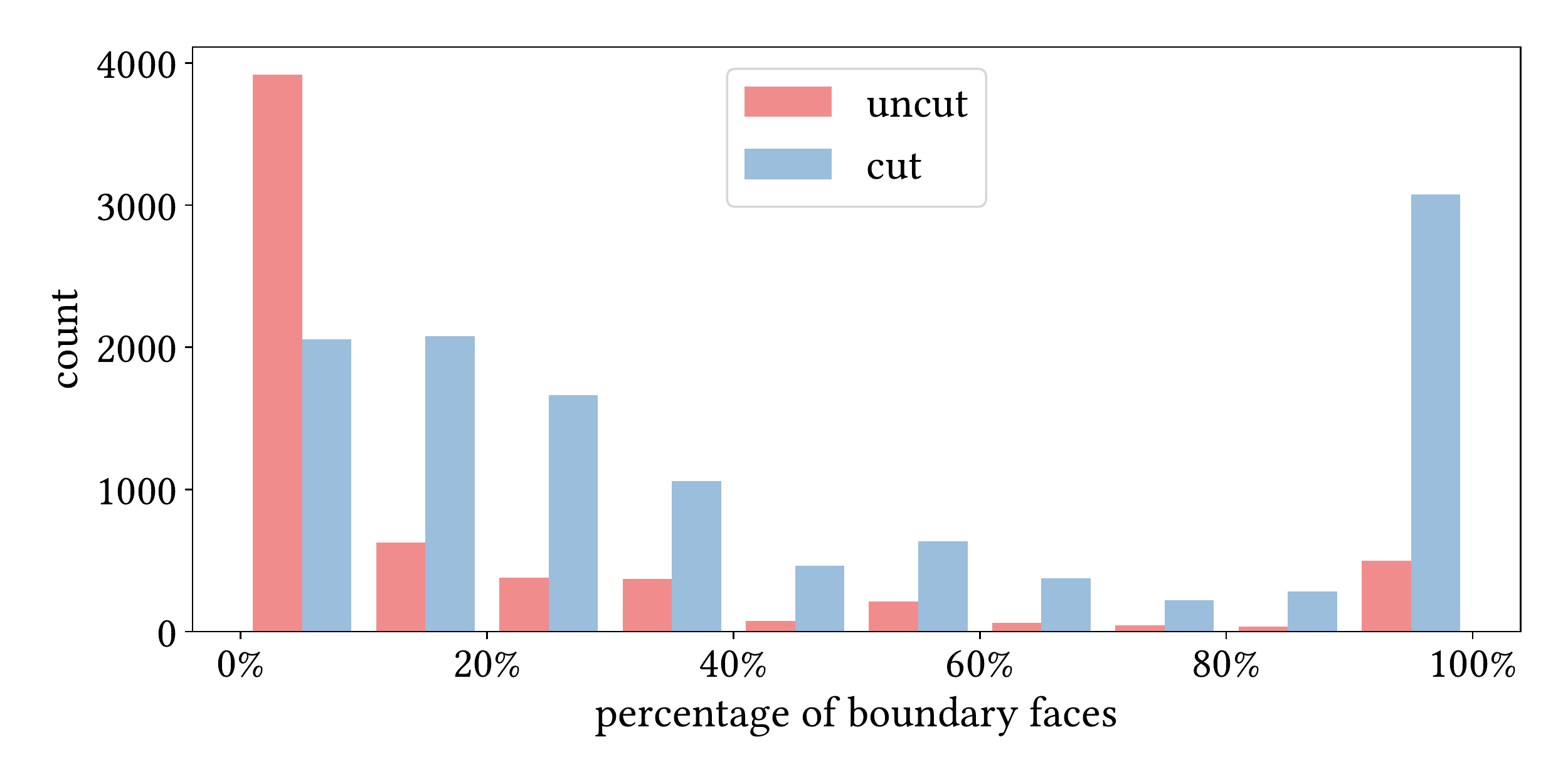}
    \includegraphics[width=\linewidth]{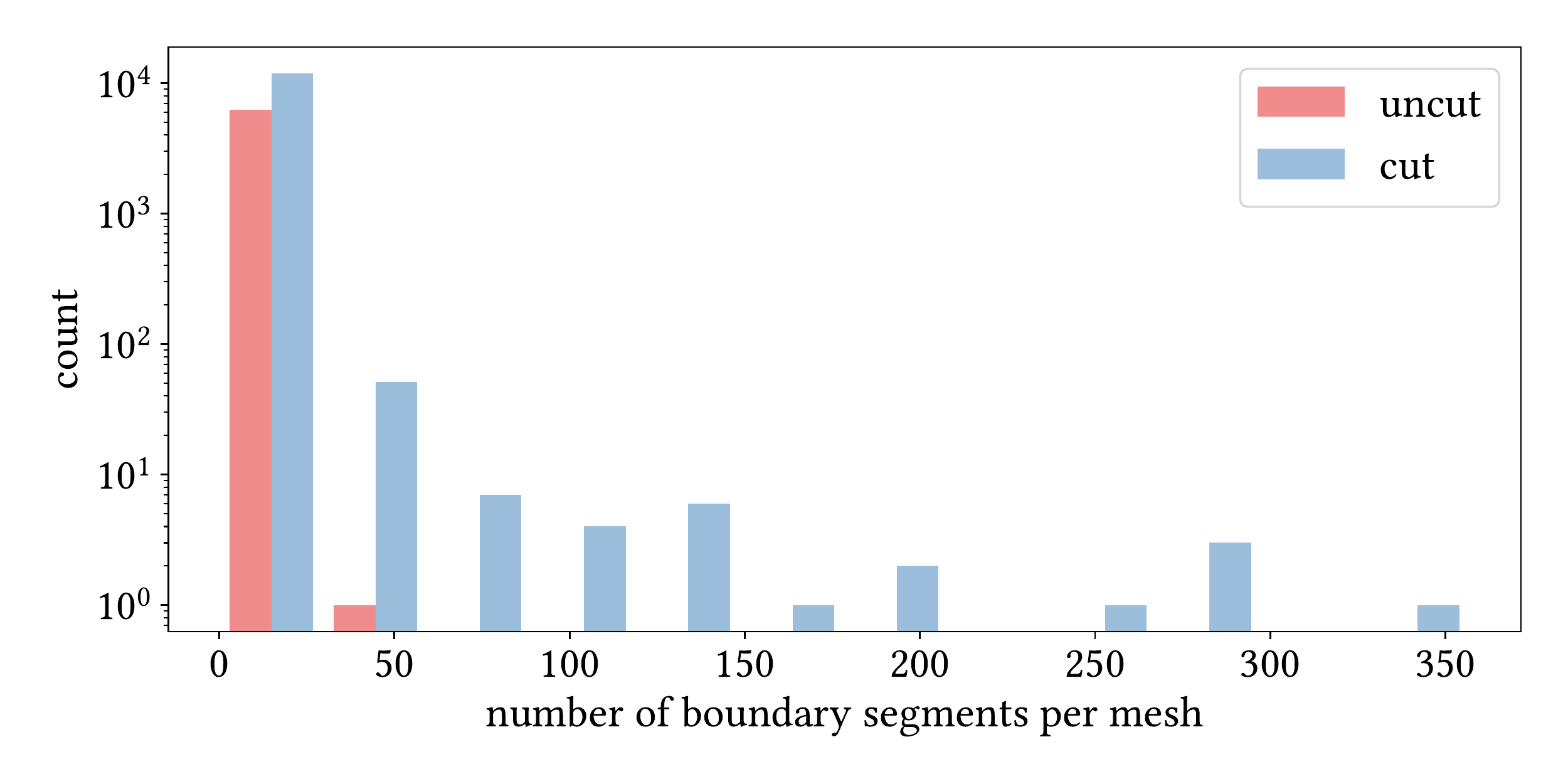}
    \includegraphics[width=\linewidth]{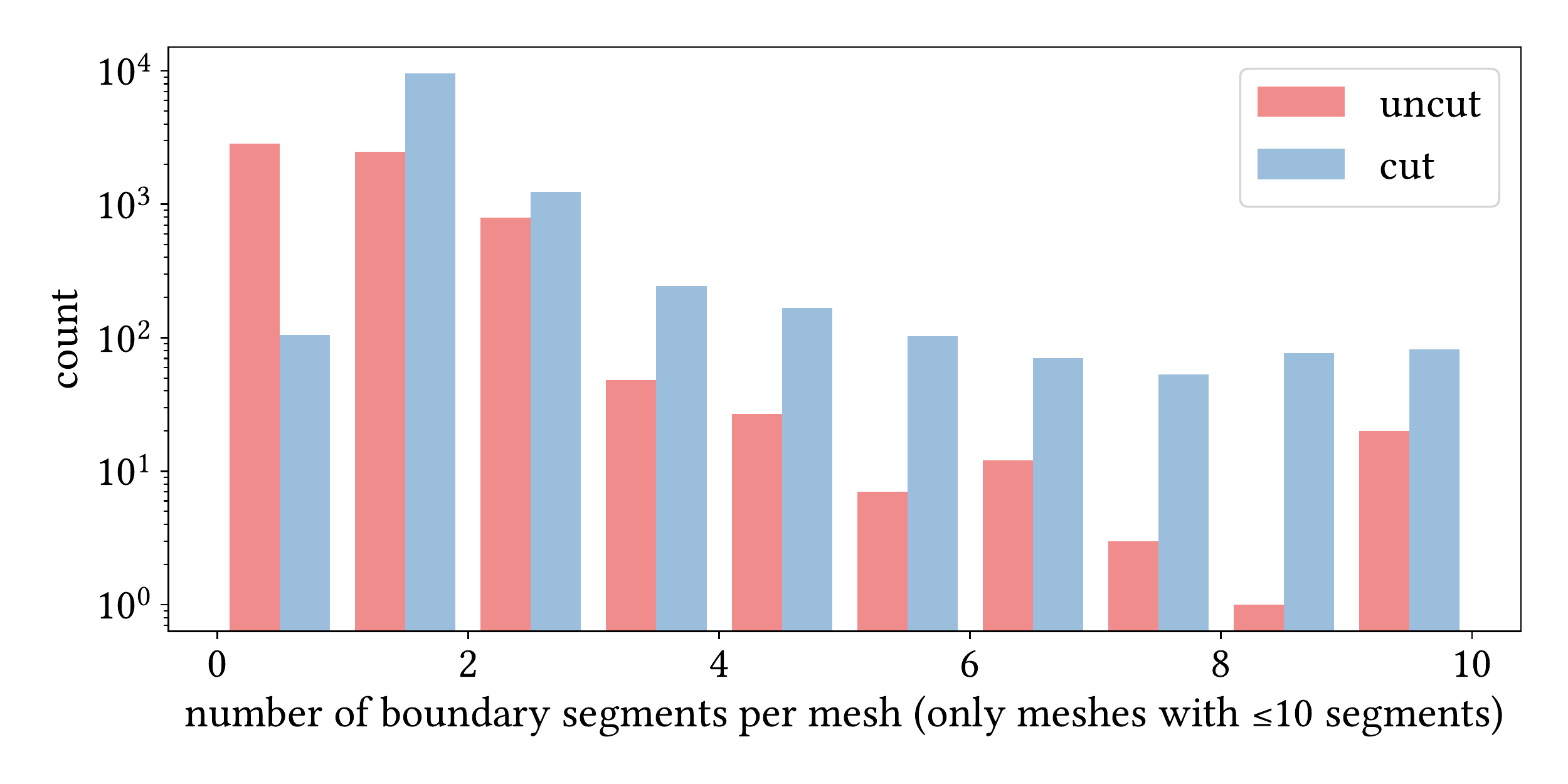}
    \caption{
    The percentage of each mesh's faces that are boundary faces (\emph{top}).
    Number of boundary segments of all meshes (\emph{middle}) as well as of
    meshes with less than or equal to 10 boundary segments (\emph{bottom}).
    }
    \label{fig:boundary}
\end{figure}

Our dataset has a variety of mesh boundaries.
The median proportion of faces lying on the boundary of the mesh is 21\%.
For cut meshes, the median is 32\% and for uncut meshes, which are more likely
to be closed, the median is 2\%.
The median number of boundary segments is 1, in the whole set as well as in the
cut and uncut parts individually (with many outliers).
For the distribution of the proportion of boundary faces and the number of
boundary segments in the dataset, see Fig.~\ref{fig:boundary}.
Fig.~\ref{fig:outlier} shows an example of an outlier with 357 boundary segments.

\begin{figure}
    \centering
    \includegraphics[width=0.49\linewidth]{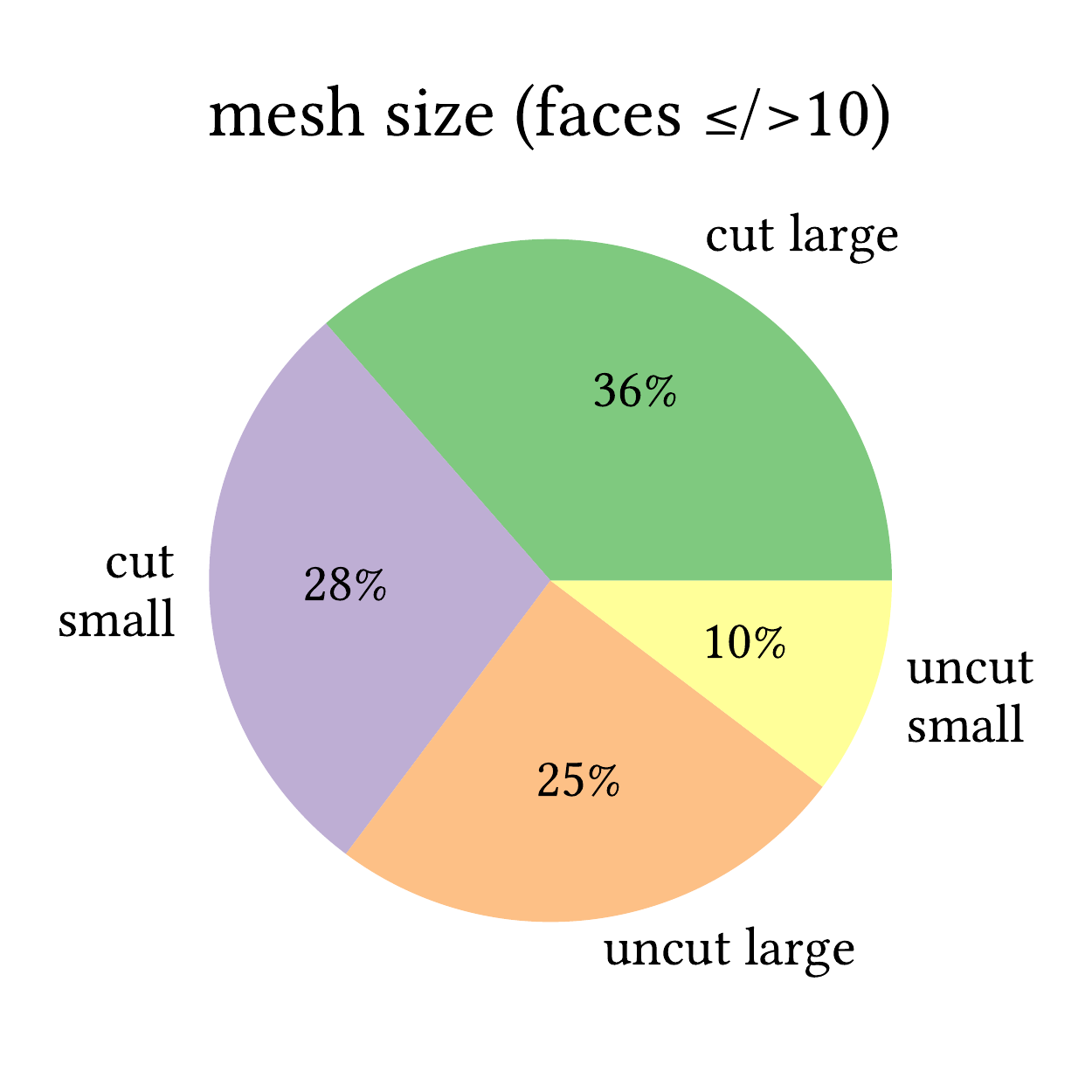}
    \includegraphics[width=0.49\linewidth]{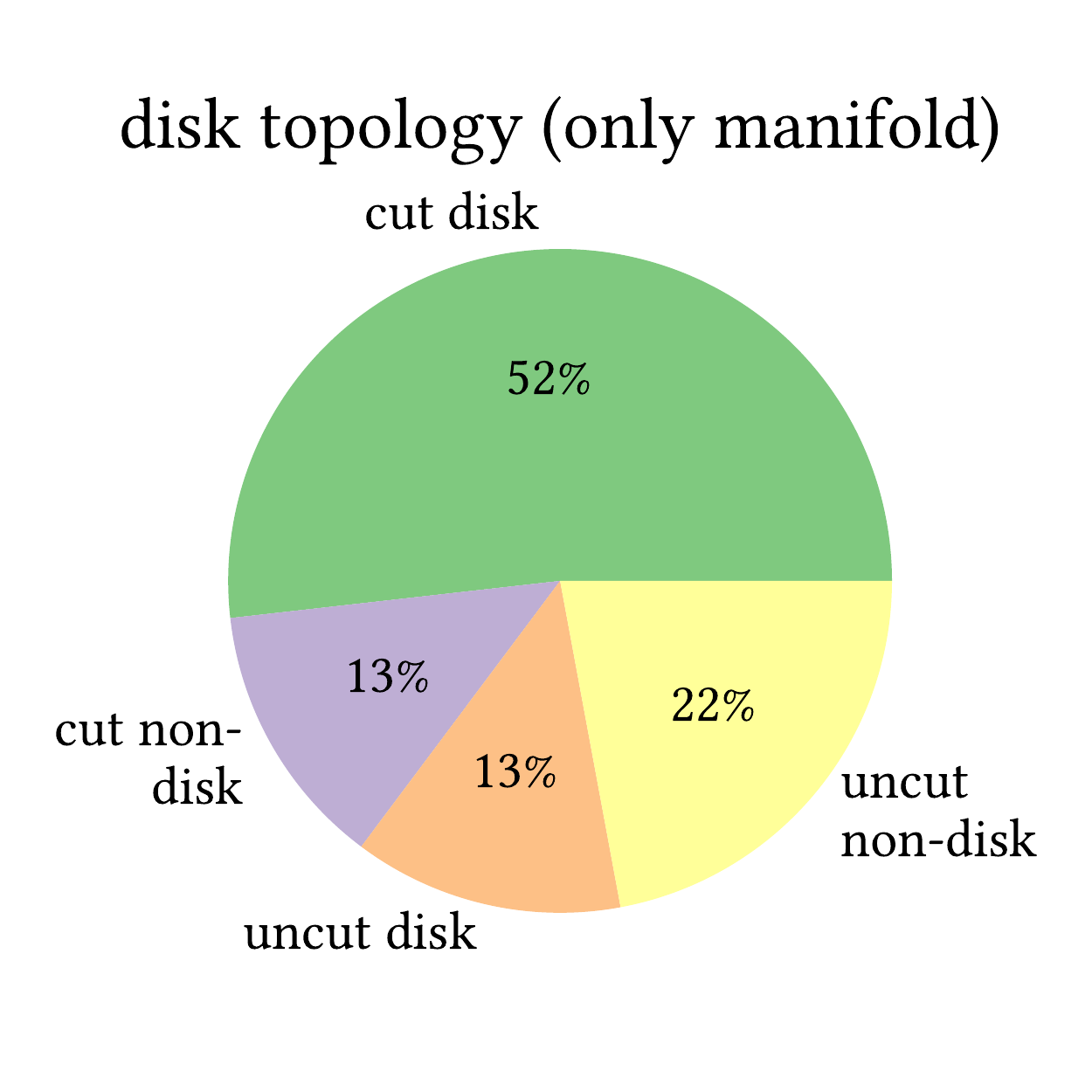}
    \includegraphics[width=0.49\linewidth]{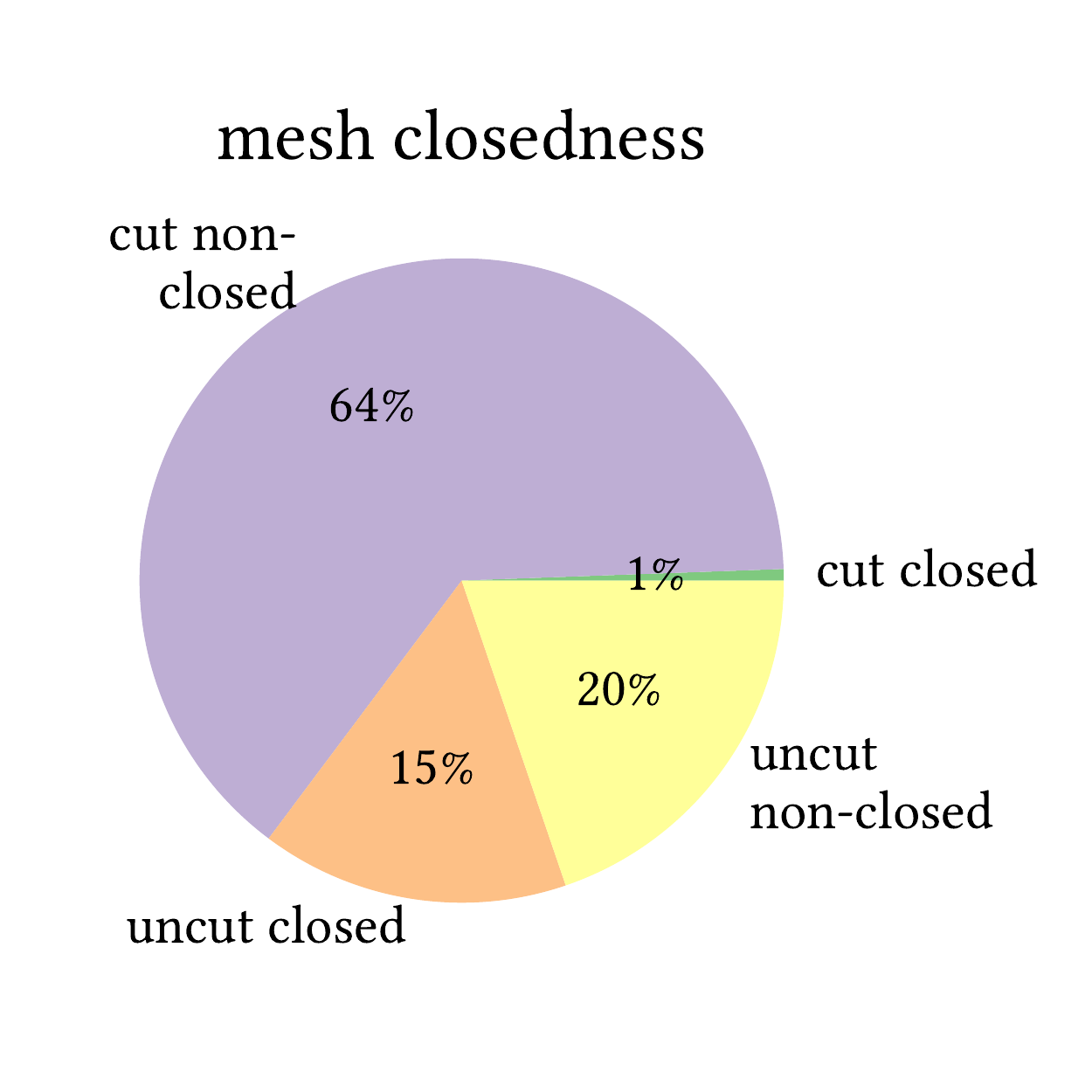}
    \includegraphics[width=0.49\linewidth]{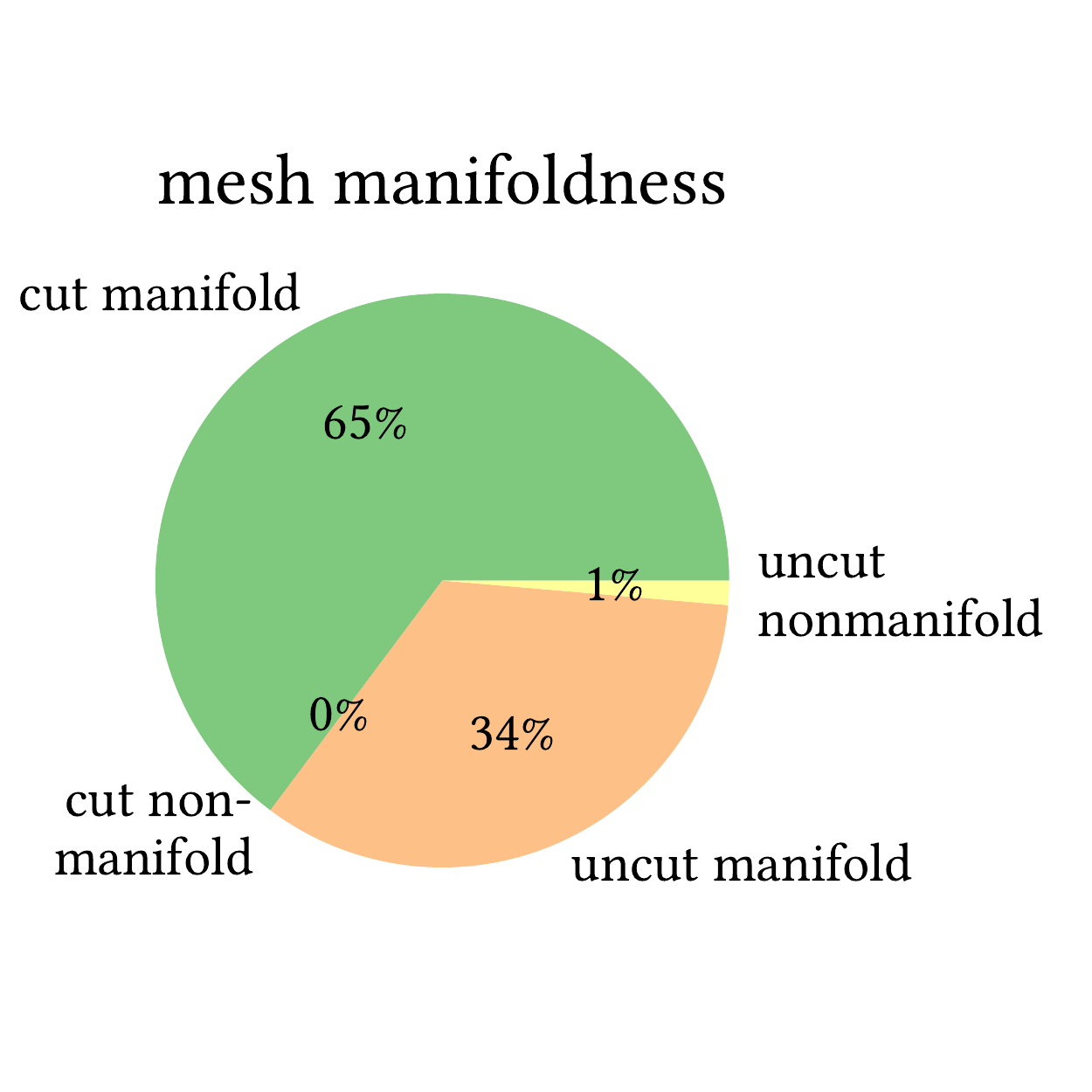}
    \caption{Pie charts showing the distribution of tags in different parts
    of the dataset.
    Our benchmark can be run on subsets of the dataset with a certain tag
    property, to accommodate methods that only work on certain kinds of inputs.}
    \label{fig:tags}
\end{figure}

As described in \S\ref{sec:dataset-tags}, we partition our dataset using four tags. Most of these tags are fairly evenly distributed across the dataset (see Fig.~\ref{fig:tags}).
We note here where this is not the case.
The vast majority of cut meshes are open.
In most cases, either the mesh has a boundary or the mesh was cut by the artist before creating a texture map. Most meshes in our dataset also are manifold. For the cut meshes, this property is expected, as all nonmanifold cut meshes are filtered out. About 1\% of our uncut meshes are nonmanifold.

\section{Benchmark}

Our benchmark provides a way to evaluate the performance of a parameterization algorithm on our dataset by measuring quantities of interest, and comparing them to the artist-provided UV map.
The benchmark is a simple program that is run on a full set of UV maps computed by the method to be benchmarked for the meshes in the dataset.

\subsection{Measures}
\label{sec:measures}

\begin{figure*}
    \centering
    \includegraphics[height=0.15\linewidth]{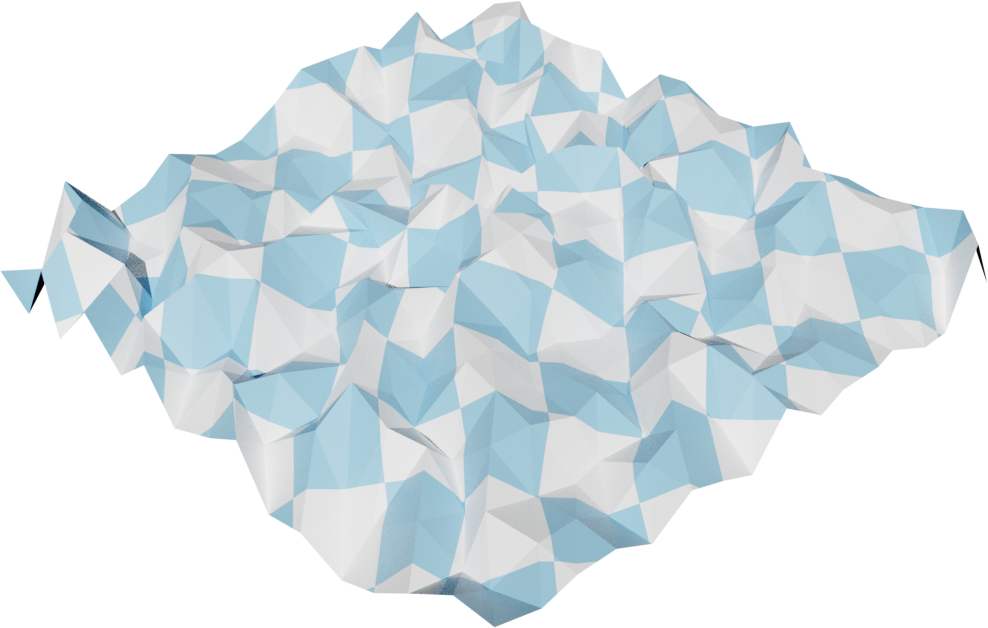}
    \includegraphics[height=0.15\linewidth]{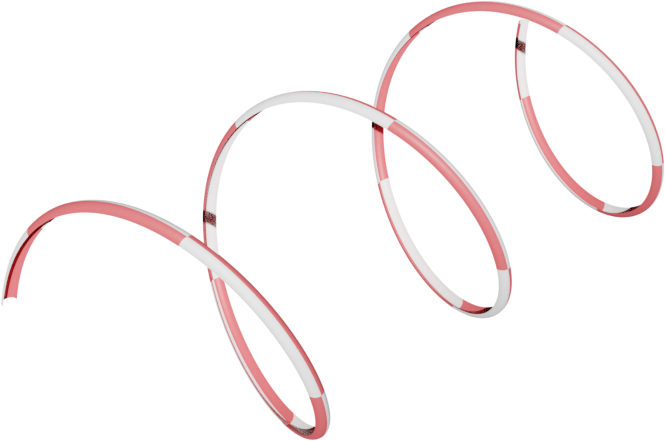}
    \includegraphics[height=0.15\linewidth]{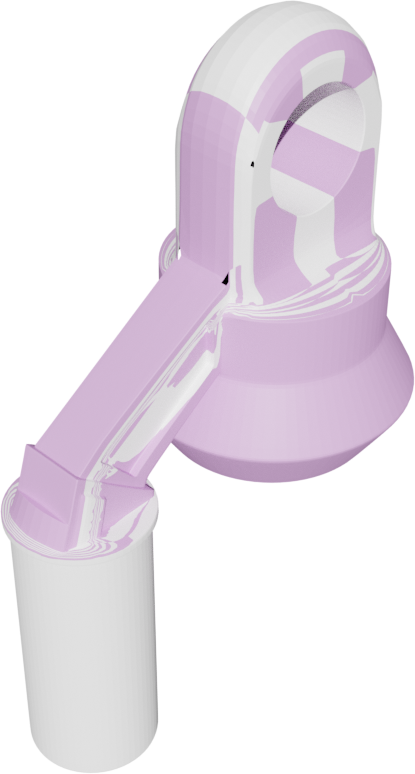}
    \includegraphics[height=0.15\linewidth]{assets/trimmed/object_159_140mm_fan_cover_0.png}
    \includegraphics[height=0.15\linewidth]{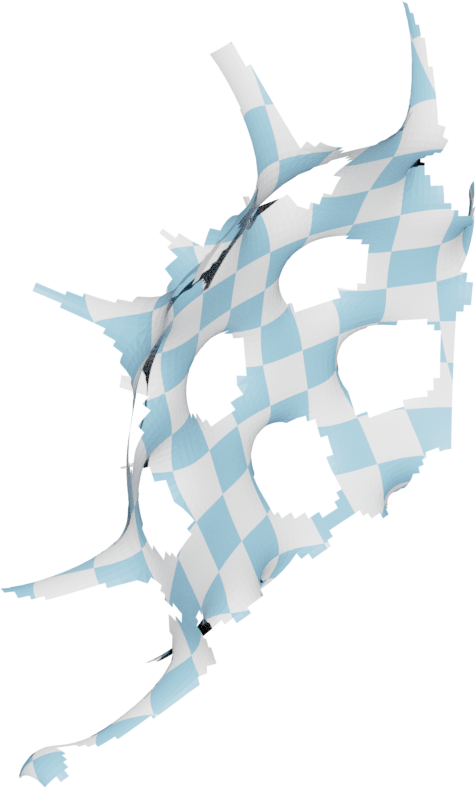}
    \includegraphics[height=0.15\linewidth]{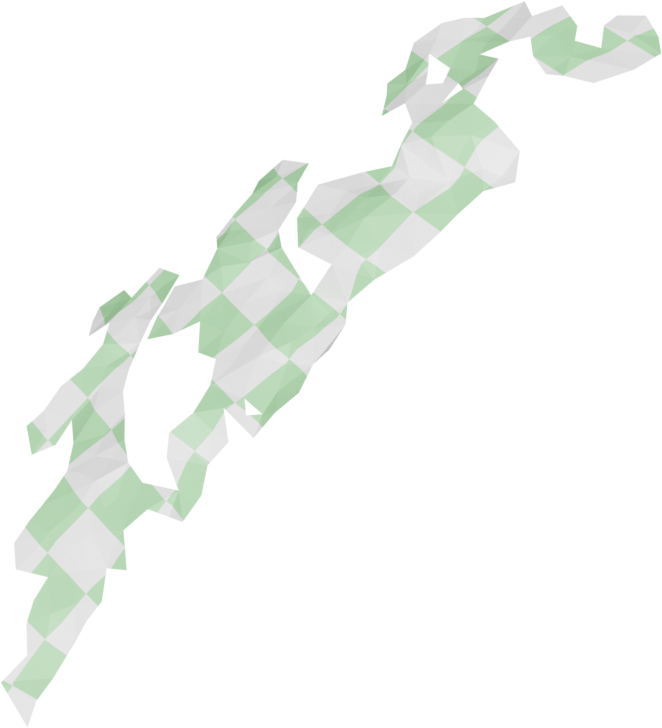}
    \caption{Examples for \emph{interesting meshes}.
    These are meshes that we pay close attention to in our benchmark.
    Instead of just computing one quantity per mesh and then aggregating them in a large plot (as is done for all meshes), we include a plot of per-triangle quantities for these meshes (Fig.~\ref{fig:pipe}).}
    \label{fig:interesting-meshes}
\end{figure*}

Our benchmark calculates and reports basic statistical information (filename, faces, vertices) about each mesh as well as the following metrics for the parameterizations provided. With the exception of resolution, all of the following metrics are calculated after scaling the mesh and UV map such that both have total area 1.

In this section, we adopt the convention that the area of a triangle $T$ is $A_T$, its UV area is $A_{UV, T}$, and the singular values of its UV map are $\sigma_{1, T}$ and $\sigma_{2, T}$.
The three angles of $T$ are $a_T$, $b_T$, and $c_T$, and the corresponding angles on its UV triangle are $\alpha_T$, $\beta_T$, and $\gamma_T$.
The surface containing all triangles is $\Omega$.

Below, we list the metrics reported by our benchmark.
For all of these metrics we compute one value per mesh, an average or an extreme, that measures some aspect of parameterization quality.
We output these metrics into a CSV file and then aggregate them into one plot per metric, containing data from all meshes in the dataset.

\paragraph*{Maximum area distortion.}
The area distortion of a triangle $T$ is
$$D_{Area, T} = \frac{A_T}{A_{UV, T}} + \frac{A_{UV, T}}{A_T} - 2.$$
This distortion is minimized at $D_{Area, T} = 0$ when $A_T = A_{UV, T}$. If either $A_T$ or $A_{UV, T}$ is $0$, the result is reported as infinity.

The \emph{maximum area distortion} is $\max_{T \in \Omega} D_{Area, T}$.
We ignore triangles $T$ for which the area on the mesh (not the UV area) is so small that our computation of a map from $T$ causes the appearance of NaNs.

\paragraph*{Average area discrepancy.}
The area discrepancy of a triangle
$T$ is
$$E_{Area, T} = |A_T - A_{UV, T}|.$$

The \emph{average area discrepancy} is
$$\frac{1}{\sum_{T \in \Omega} A_T} \sum_{T \in \Omega} E_{Area, T}.$$
This discrepancy is minimized with
$E_{Area} = 0$
when all triangles have the same proportion of area in UV space as they do in
mesh space.

\paragraph*{Minimum and maximum singular values.}
Each triangle's UV map is a 2D affine map and thus has two singular values.
These indicate the amount of stretch the triangle has undergone in orthogonal directions. Both the \emph{minimum and maximum singular values} across all triangles in the mesh are reported. These singular values need not come from the same triangle.

\paragraph*{Percentage of flipped triangles.}
A triangle is \emph{flipped} if its orientation in the mesh is opposite 
to
its orientation in the UV map, i.e., if the Jacobian $J$ of the linear transformation
from mesh to UV triangle has a negative determinant.
We report the percentage of triangles in the mesh that are flipped under the parameterization.
Since a mesh with $x$\% flipped triangles can easily be converted into a mesh with $100-x$\% flipped triangles by mirroring, we report $\min{(x, 100-x)}$\%.

\paragraph*{Maximum angle distortion.}
The angle distortion of a triangle $T$ is
$$D_{Angle, T} = \frac{\sigma_{1, T}}{\sigma_{2, T}} + \frac{\sigma_{2, T}}{\sigma_{1, T}} - 2.$$

This distortion is minimized at $D_{Angle, T} = 0$ when $\sigma_1 = \sigma_2$, i.e., when the triangle has been stretched equally in all directions, conserving angles.
If either $\sigma_1$ or $\sigma_2$ is $0$, the result is reported as infinity. The \emph{maximum angle distortion} is $\max_{T \in \Omega} D_{Angle, T}$.

\paragraph*{Average angle discrepancy.}
The angle discrepancy of a triangle
$T$ is
$$E_{Angle, T} = |a_T - \alpha_T| + |b_T - \beta_T| + |c_T - \gamma_T|.$$
The \emph{average angle discrepancy} is
$$\frac{1}{\sum_{T \in \Omega} A_T} \sum_{T \in \Omega} E_{Angle, T} \cdot A_T.$$
Any collapsed triangles (i.e., those with UV area equal to 0) are assigned the
maximum discrepancy of
$2\pi$. 
This discrepancy is minimized with
$E_{Angle} = 0$
when all triangles have mesh angles equal to their UV angles.
It has a maximum value of $2\pi$.

\paragraph*{Symmetric Dirichlet energy.}
The symmetric Dirichlet energy of a triangle $T$ is given by \cite{Smith2015}
$$E_{Dir, T} = \frac12 \left( \sigma_{1,T}^2 + \sigma_{2,T}^2 +
\frac{1}{\sigma_{1,T}^2} + \frac{1}{\sigma_{2,T}^2} \right),$$
where the
$\sigma_{i,T}$
are the singular values of the Jacobian on
$T$.
The symmetric Dirichlet energy of the whole mesh is
$$E_{Dir} = \sum_{T \in \Omega} A_T E_{Dir,T}$$
This energy is sensitive to both nonrigid transformations in general, as well as
triangle collapses specifically, as maps that collapse triangles have at least
one singular value equal to zero.

\paragraph*{Resolution.}
To compute this quantity, the area of the original mesh is not rescaled to equal 1.
The UV map is repositioned to fit into the unit square.

For a triangle $T$ with singular values $\sigma_{1, T}$ and $\sigma_{2, T}$, the resolution required for the UV map to achieve a resolution of 1 on the given triangle is
$$R_{T} = \max{\left(\frac{1}{\sigma_{1, T}}, \frac{1}{\sigma_{2, T}}\right)}.$$
If a triangle is stretched by $\sigma$ in a direction, then we need a texture with a resolution of at least $\frac{1}{\sigma}$ to display this triangle in a hypothetical (scaled) rendering. The minimum resolution required is then the maximum of these resolutions across all triangles.

\paragraph*{Artist correlation.}
This metric measures whether the parameterization algorithm chooses to distort
the triangles in the same way as the artist, i.e., whether the same triangles are
blown up or shrunk down.
Let
$\sigma_{UV,1,T}, \sigma_{UV,2,T}$
be the singular values of the Jacobian of the parametrization to be benchmarked
on triangle
$T$,
and let
$\sigma_{ART,1,T}, \sigma_{ART,2,T}$
be the respective singular values of the artist-derived parameterization.
We define the averages
$\mu$,
the covariance
$c$,
and the standard deviation
$s$
as
\begin{equation*}\begin{split}
\mu_{\alpha} &= \frac{1}{2 \sum_{T \in \Omega} A_T}
\sum_{T \in \Omega} A_T \sum_i \sigma_{\alpha,i,T}\\
c_{\alpha,\beta} &= \frac{1}{2 \sum_{T \in \Omega} A_T}
\sum_{T \in \Omega} A_T \sum_i
(\sigma_{\alpha,i,T} - \mu_{\alpha})(\sigma_{\beta,i,T} - \mu_{\beta}) \\
s_{\alpha} &= \sqrt{c_{\alpha,\alpha}},
\end{split}\end{equation*}
where
$A_T$
is the area of triangle
$T$
and the sum is over all triangles.
The artist correlation (where smaller values imply more correlation) is then
defined as
$$\left|
\frac{c_{UV, \, ART}}{s_{UV, \, UV} \; s_{ART, \, ART}}
- 1 \right|.$$

This metric is similar to Pearson's correlation coefficient \cite{pearsoncorrelation},
but does not really measure a correlation in the statistical sense, since we
cannot assume that the relevant quantities follow a normal distribution.
It is important to note that this metric is \emph{not} a measure of
parameterization geometric quality.
The artist correlation merely measures whether the tested parameterization
method has the same priorities as the artist when choosing on which triangles
to place distortion.

\paragraph*{Remeshed.}
There are methods methods (such as CEPS \cite{discrete-conformal}) that employ retriangulation for their parameterizations.
To detect this change, we check if any of the following holds:
\begin{itemize}
    \item The number of vertices differs from the original mesh.
    \item The number of faces differs from the original mesh.
    \item Any face references different vertices.
    \item The difference between the position of a vertex in the original mesh and a vertex in the provided mesh is greater than $10^{-5}$ for any vertex.
\end{itemize}
If so, we mark the parameterization as \emph{remeshed} and do not compute per-triangle comparisons with the artist (as the triangles are no longer comparable).

\paragraph*{Mesh cut length.}
For methods run on the uncut dataset, we include a measure of the \emph{mesh cut length}.
This is the length of cuts on the mesh, with the mesh area rescaled to a total of 1. For meshes that have boundaries initially, the length of these boundaries is not counted and only the length of new cuts is reported.

\paragraph*{Artist mesh cut length match.}
For methods run on the uncut dataset, we report an
\emph{artist's match of mesh cut length}.
This value measures how a parameterization's cut length compares to the artist's
parameterization in mesh space. The cut length is computed on the mesh itself.
For a cut length $C$ and artist cut length $C_{ART}$, we report
$$M_{Cut, Mesh} = \max{(0, C - C_{ART})}.$$
This measure is $0$ if the method yields a smaller cut length than the
artist's parameterization, and it equals the excess cut length otherwise.
This metric accounts for the artist's intent by being nonzero where a method
exceeds the artist's intended boundary length, and by not giving a method
credit where the artist intended to have a long cut.

\paragraph*{Interesting examples.}
Separately from the metrics where we calculate one value per mesh, we also output a separate CSV file with the singular value pairs for every triangle in a subset of ``interesting'' meshes. Some of these interesting meshes are hand-selected \emph{a priori}, due to properties such as high genus, small values, developability, or boundary triangle percentage; examples are shown in Fig.~\ref{fig:interesting-meshes}. In addition to these meshes, we also report the singular values for the meshes with the highest values for
artist correlation,
average area and angle discrepancy, and percentage flipped triangles.

\subsection{Implementation}

Our benchmark is implemented in Python 3.
We use libigl's \cite{libigl} Python bindings to import the meshes and compute
the metrics.
The benchmark is parallelized for efficiency.
The code is included in the supplemental material.

We output empty rows in the CSV file, with only filename and the number of faces
and vertices, for any meshes in the dataset that the method did not
parameterize, any meshes where there are NaN texture coordinates,
any parameterization with area smaller than $10^{-8}$, or for any mesh where
unexpected errors occur during computation.

While many parameterization methods make theoretical guarantees, they often do
not hold under floating point arithmetic \cite{progressive-embedding}.
The computations in our benchmark are similarly subject to floating point
errors.
These errors cannot be eliminated entirely and have an effect on the resulting
metrics computed, however this effect is not observed to be large.

\section{Case Study}
\label{sec:casestudy}

\begin{figure}
    \centering
    \includegraphics[width=\linewidth]{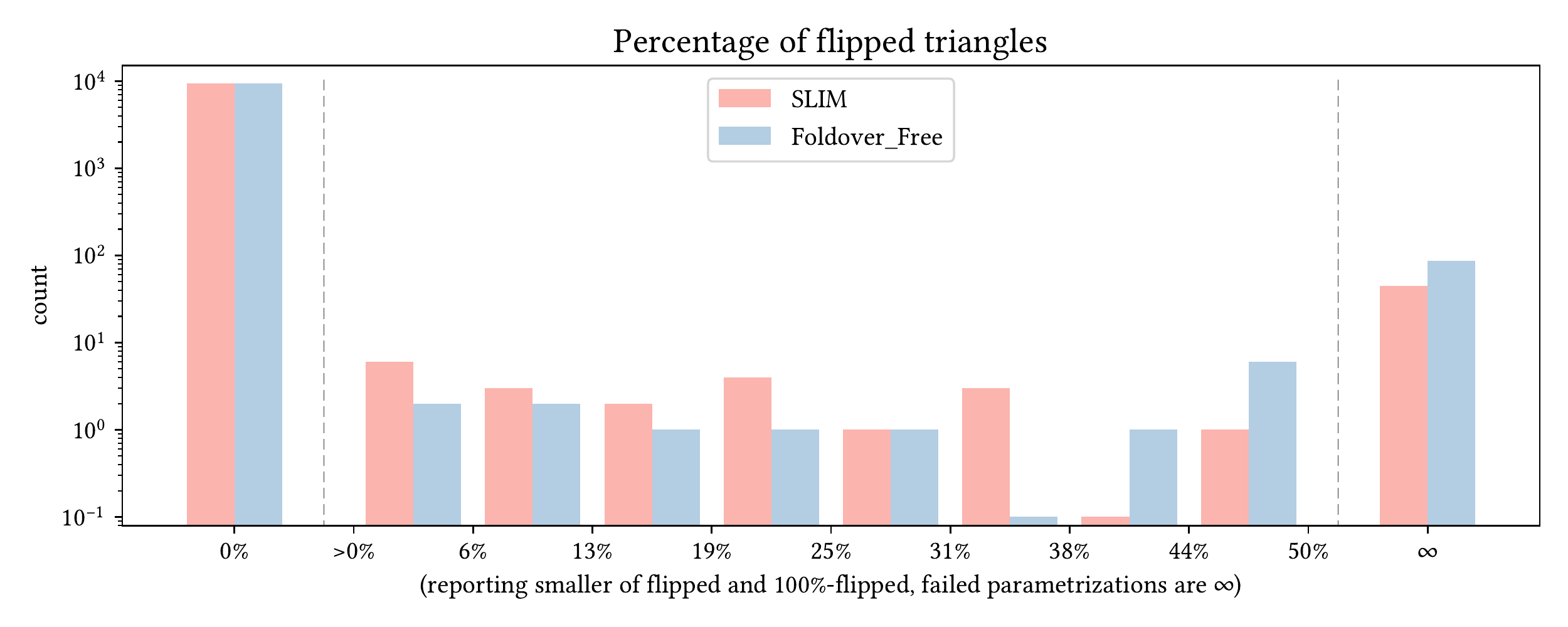}
    \caption{The percentage of flipped triangles in the UV maps generated by SLIM \cite{slim} and the Foldover-Free method \cite{foldover-free}.
    Both methods theoretically guarantee flip-free maps, but this guarantee does
    not always translate into practice due to initialization failures,
    floating point errors, or method timeouts.
    }
    \label{fig:flipped-comp}
\end{figure}

To showcase our dataset and benchmark, we
ran five different parameterization methods on our dataset and used our
benchmark to analyze the results.
The methods are a naive implementation Tutte's method \cite{tutte},
Scalable Locally Injective Mapping (SLIM) \cite{slim}, OptCuts \cite{optcuts},
Foldover-Free Maps in 50 Lines of Code (Foldover-Free)
\cite{foldover-free} (initialized, at the authors' recommendation, with
Least-Squares Conformal Mapping \cite{lscm}), and Conformal Equivalence for
Polyhedral Surfaces (CEPS)
\cite{discrete-conformal}.
All figures referenced in this section were generated directly by our
benchmarking tool. In running these parameterization methods, we used a timeout
of 5 minutes for CEPS, 5 minutes for OptCuts, 5 minutes for SLIM, and 10 minutes
for Foldover-Free.
If the method reached this timeout without finishing a mesh's
parameterizion, we cancelled the operation and did not produce an output
parameterization for that mesh.

Many parameterization algorithms specifically aim to produce flip-free
parameterizations, where no triangle has been inverted.
SLIM achieves this by initializing with a flip-free parameterization and then
performing an optimization that is guaranteed to not introduce triangle flips.
Foldover-Free can recover from flipped triangles during the initialization.
In practice, both methods produce flip-free parameterizations for most meshes
in our dataset (see Fig.~\ref{fig:flipped-comp}).
There is, however, still a significant number of meshes with flipped triangles
in the output of both methods (around 4\%).
Both methods theoretically guarantee flip-free results.
These guarantees to not always translate into practice.
This is because such theoretical guarantees make assumptions that do not always
hold in practice:
errors due to floating point arithmetic, an
execution time limit, or failure to initialize the method with a valid
state.

\begin{figure}
    \centering
    \includegraphics[width=\linewidth]{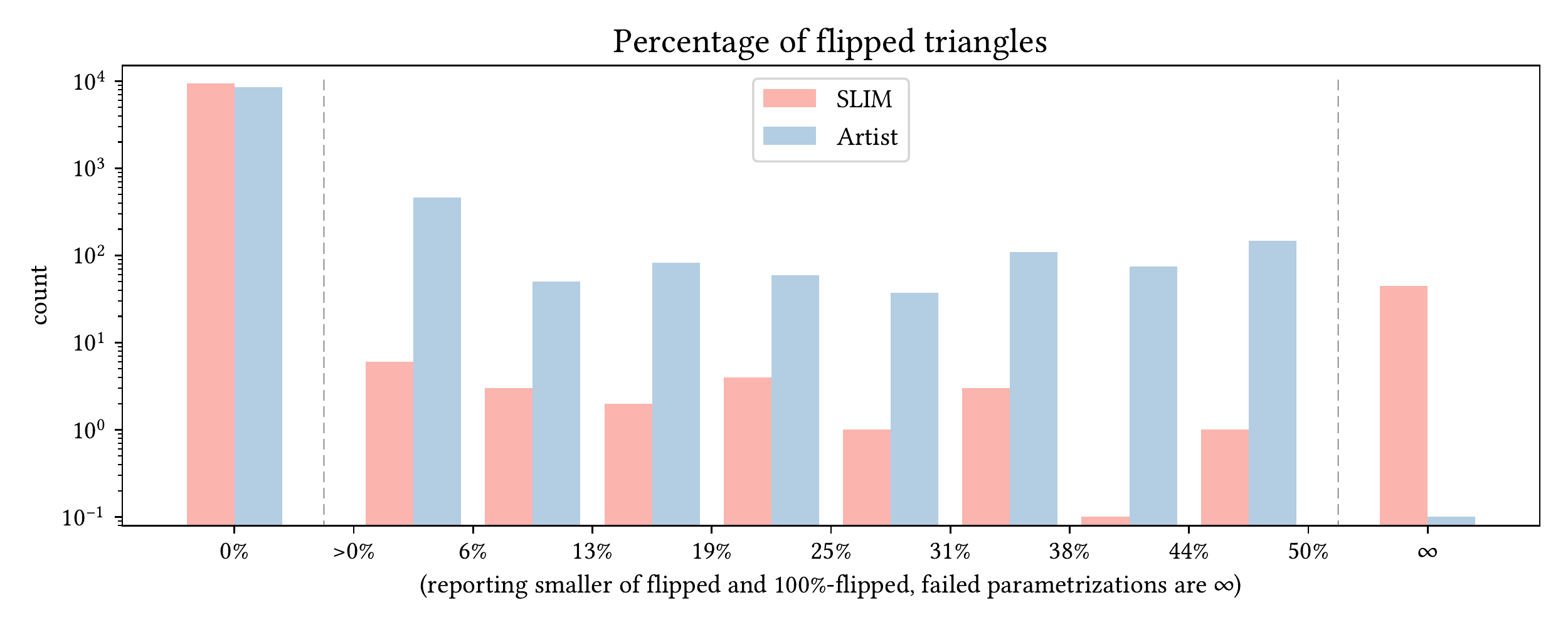}
    \caption{
    When the parameterization method is successful, SLIM \cite{slim} produces
    fewer parameterizations with flipped triangles than the artists produced
    themselves.
    }
    \label{fig:flipped-slim}
\end{figure}

Fig.~\ref{fig:flipped-slim} shows that SLIM produces, in general, fewer
parameterizations with flipped triangles (when it can successfully produce a
parameterization) than the artists themselves.
This can indicate that SLIM is better than the artists at producing flip-free
maps, or it can indicate that artists do not always value flip-free maps enough
in practice to prioritize the property (see also Fig.~\ref{fig:cowhide}).

\begin{figure}
    \centering
    \includegraphics[width=\linewidth]{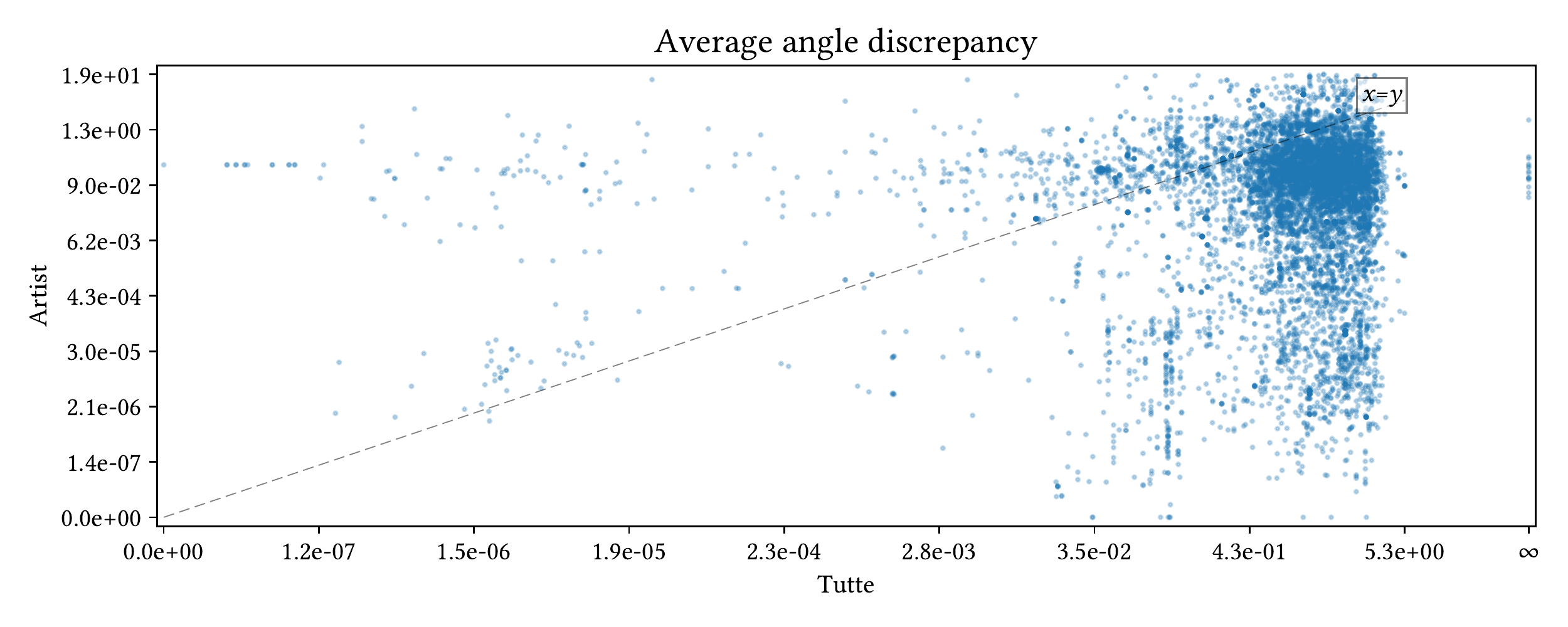}
    \includegraphics[width=\linewidth]{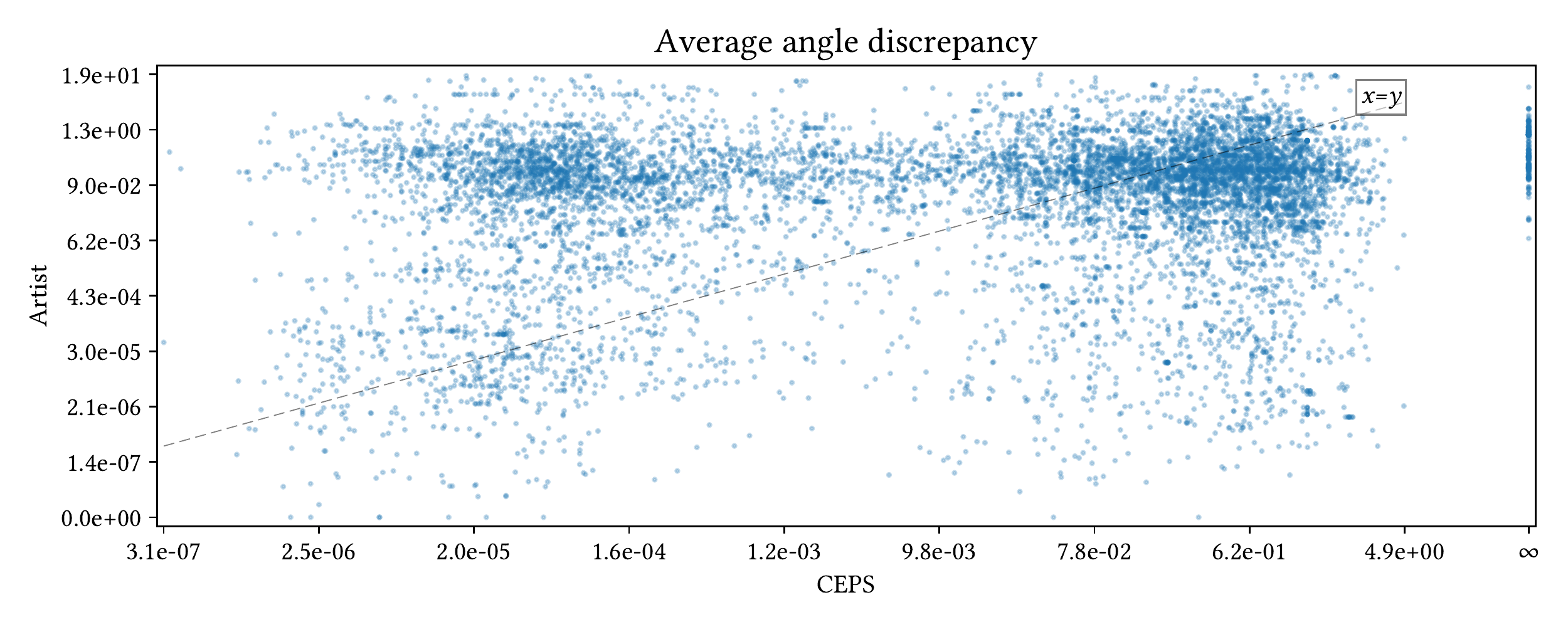}
    \caption{Comparing the
    average angle discrepancy
    of a naive implementation of Tutte's method and CEPS \cite{discrete-conformal} to the provided UV maps.
    While the provided UV maps tend to have lower discrepancy than Tutte's, CEPS produces lower discrepancy than the artist.}
    \label{fig:area-discrepancy-tutte-ceps}
\end{figure}

CEPS retriangulates meshes to optimize for angle distortion
\cite{discrete-conformal}.
Comparing CEPS and our naive implementation of Tutte's method \cite{tutte} in
a scatter plot, we can see that while Tutte produces more discrepancy than the
artist, CEPS outperforms artist texture maps on this metric
(Fig.~\ref{fig:area-discrepancy-tutte-ceps}).

\begin{figure}
    \centering
    \includegraphics[width=\linewidth]{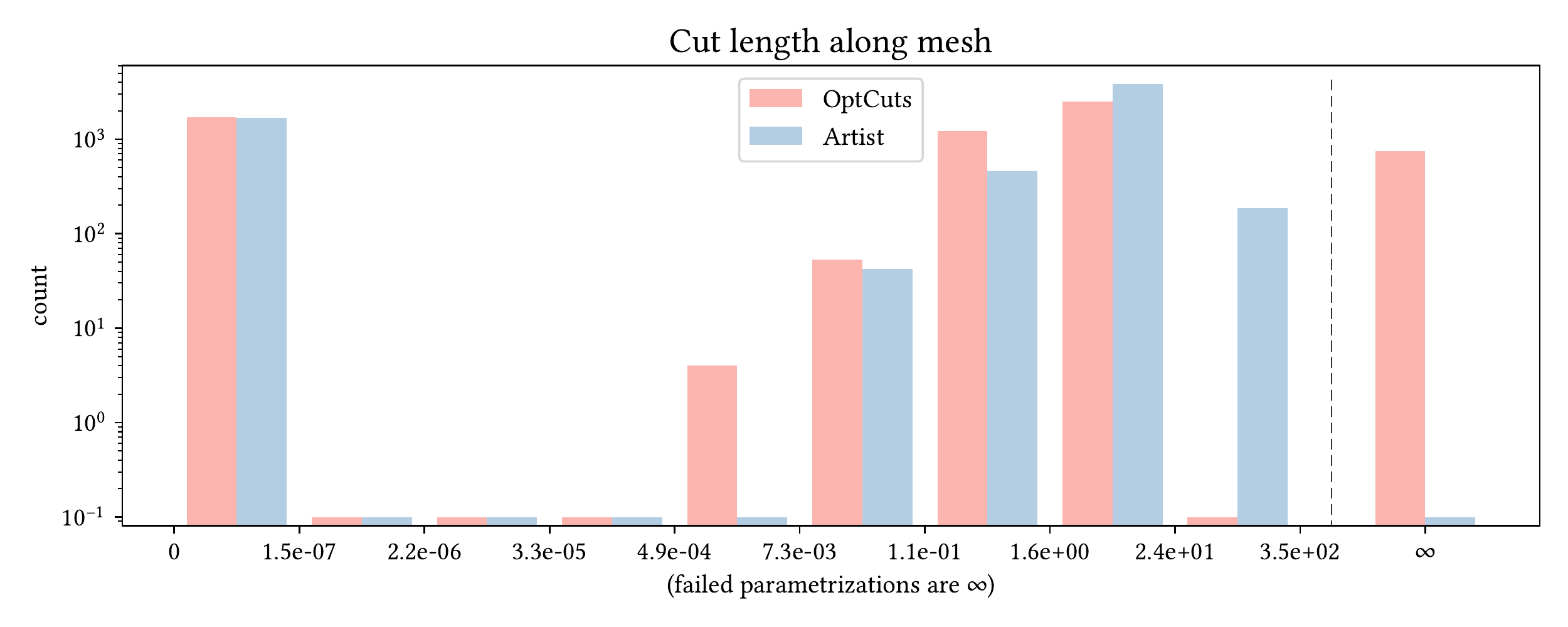}
    \caption{A histogram of the length of UV seams for OptCuts \cite{optcuts} and the provided UV maps.
    Where OptCuts does not fail, it seems to match (and even improve on) the cut length of the artist.}
    \label{fig:cut-length-optcuts}
\end{figure}

\begin{figure}
    \centering
    \includegraphics[width=\linewidth]{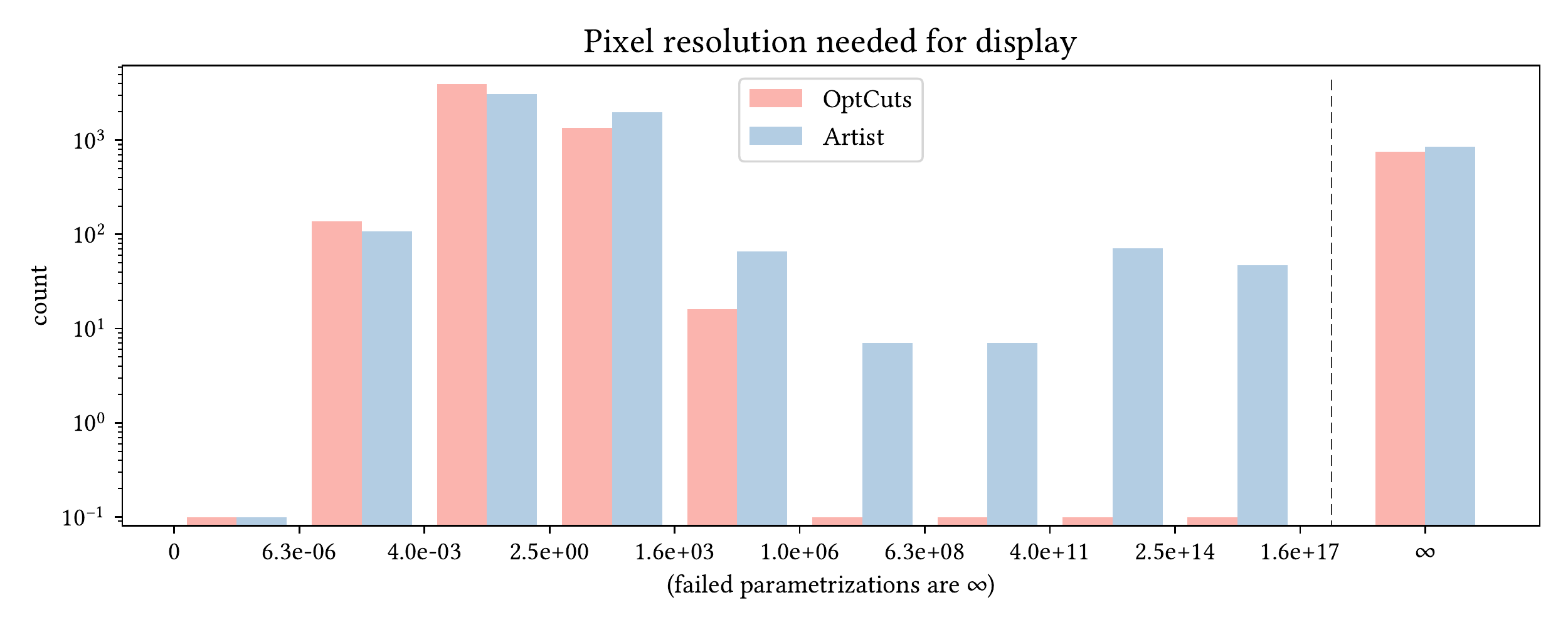}
    \caption{The relative texture resolution required to texture an image with the UV map computed by OptCuts \cite{optcuts} and the provided UV map.
    OptCuts consistently outperforms the artist.
    This does not always not matter in practice, since the artist can deliberately choose to not care about triangles they deem irrelevant, while OptCuts accounts for all triangles.}
    \label{fig:resolution-optcuts}
\end{figure}

OptCuts \cite{optcuts} divides meshes into patches before parameterization,
optimizing the cut lengths and splitting into charts until a bounded level of
distortion is achieved.
Unlike other methods in this survey, this means that OptCuts can parameterize
meshes from the uncut portion of our dataset, a task that is highly relevant
in practice.
Our benchmark can evaluate the cutting task performance specifically.
Fig.~\ref{fig:cut-length-optcuts} shows that OptCuts, in general, cuts less than
the artist.
In Fig.~\ref{fig:resolution-optcuts} we can see that OptCuts produces
UV maps that do not require as high resolutions to display textures as the
artist-generated UV maps.
We speculate that is because the artist knows which parts of the surface are
important (and thus should have a low required pixel resolution) and which parts
are irrelevant, and can use this information to inform the construction of the
UV map.

\begin{figure}
    \centering
    \includegraphics[width=\linewidth]{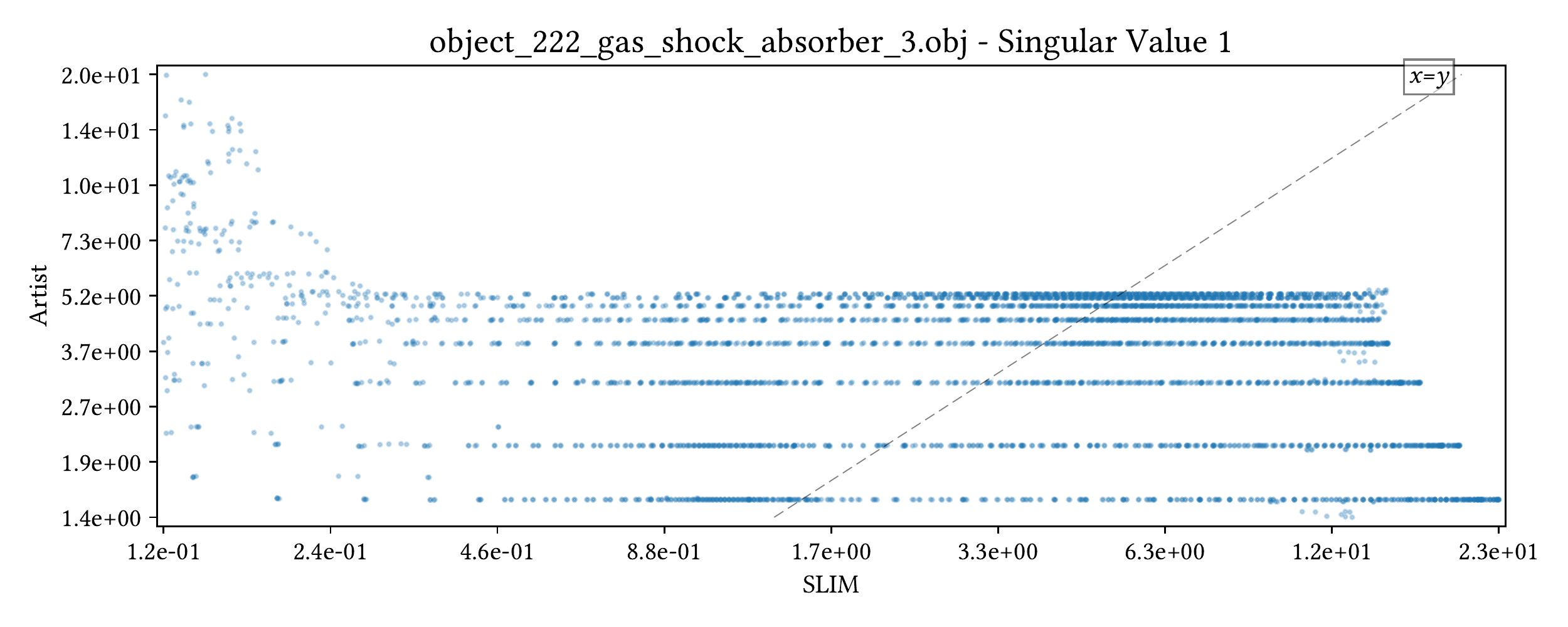}
    \includegraphics[height=0.3\linewidth]{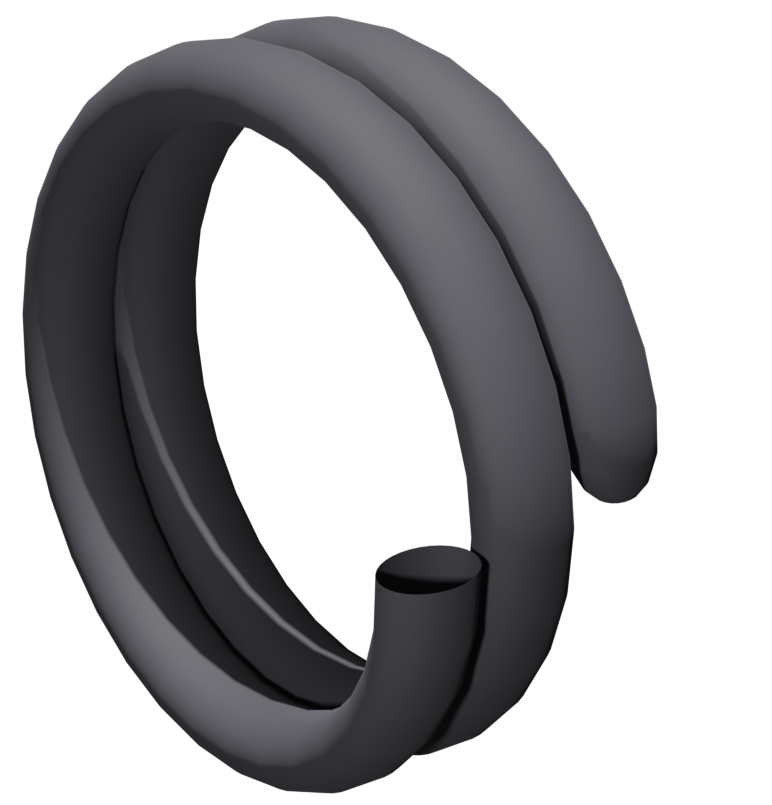}
    \includegraphics[height=0.28\linewidth]{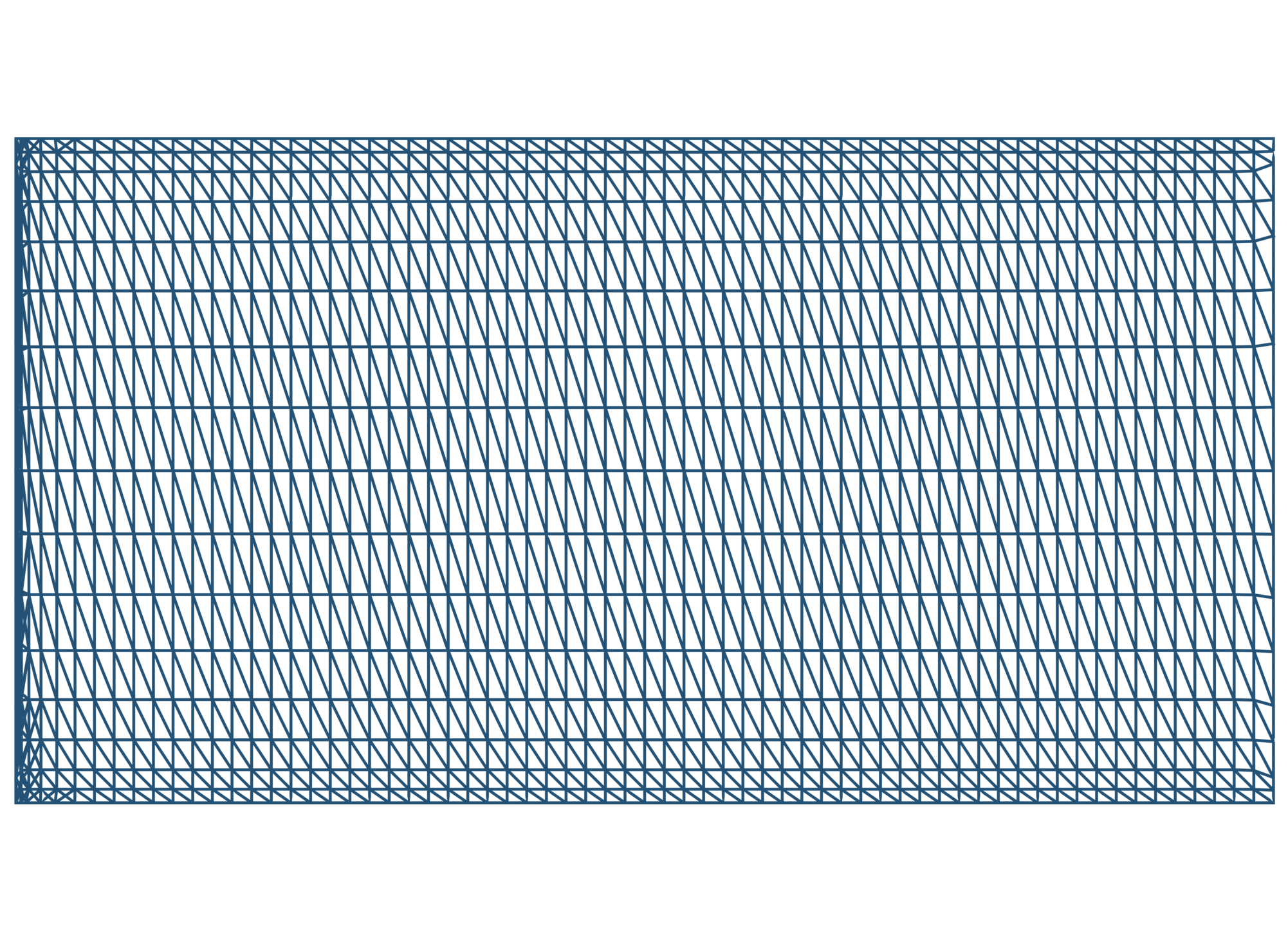}
    \hspace{5pt}
    \includegraphics[height=0.3\linewidth]{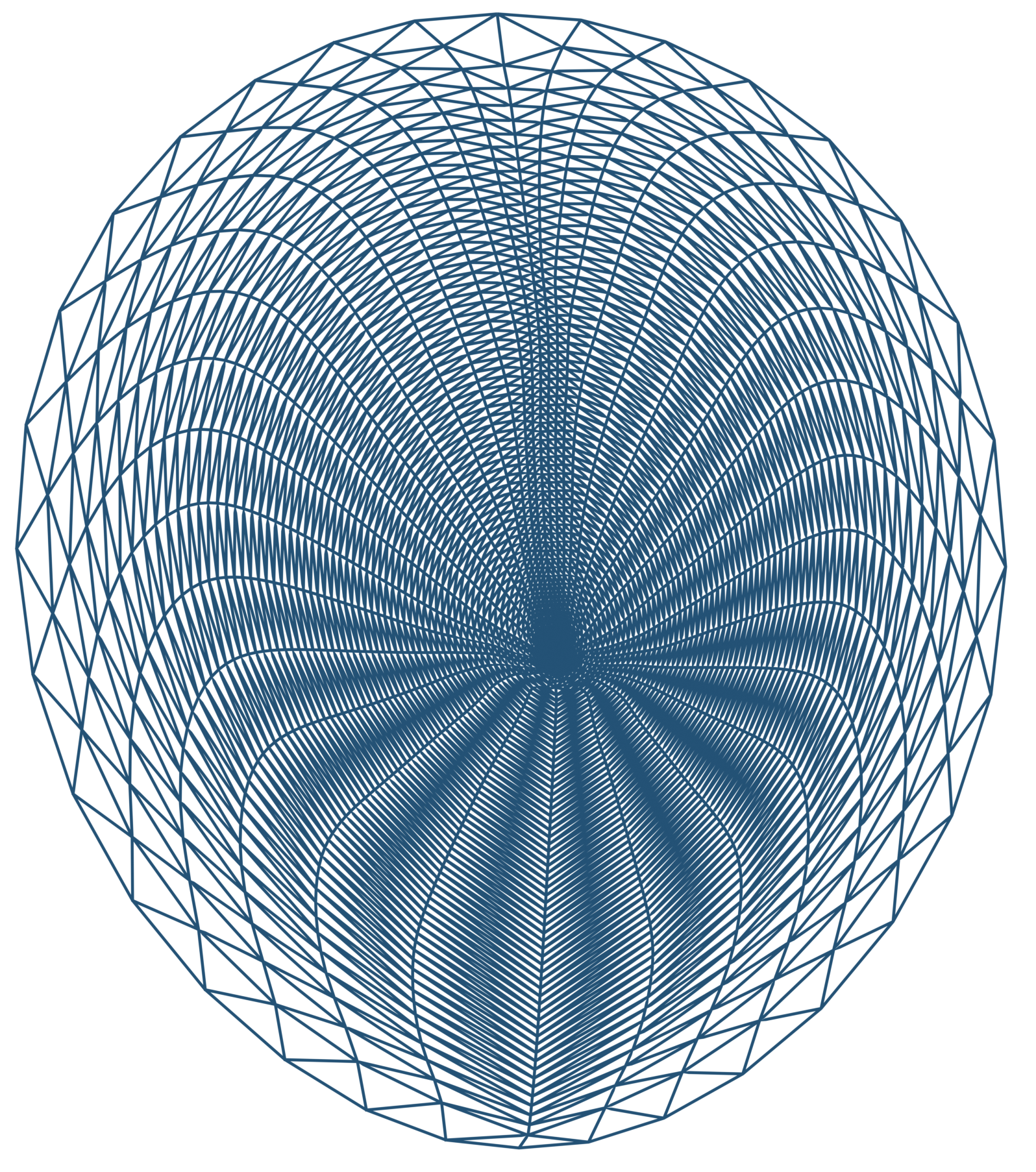}
    \caption{The smaller singular value of the Jacobian of each face in an \emph{interesting mesh} (\emph{top}) for SLIM \cite{slim} and the provided UV map.
    The bottom row displays the mesh (\emph{bottom left}), the provided UV map (\emph{bottom center}), and SLIM's UV map (\emph{bottom right}).
    A lot of the artist's singular values cluster into discrete bins, which is not the case for SLIM.
    This is because the artist exploits the cylindrical nature of the pipe, makes a cut along height of the (undeformed) cylinder, and then flattens the surface as if it were a cylinder.}
    \label{fig:pipe}
\end{figure}

Fig.~\ref{fig:pipe} shows an example from the ``interesting meshes'' part of our
benchmark.
Here, the artist's UV map features highly discrete singular values, while SLIM
produces a more continuous result.
We hypothesize that this is due to the artist's semantic knowledge:
since they knew that the pipe is a deformed cylinder, they simply made a cut
along the length of the cylinder and made the UV map a simple rectangle using
semantic information.
SLIM does not use this additional information.
It is an automatic paramaterization method without the ability to cut the mesh,
and so just flattened the mesh using the existing boundary trying to reduce
global distortion.

\begin{figure}
    \centering
    \includegraphics[width=\linewidth]{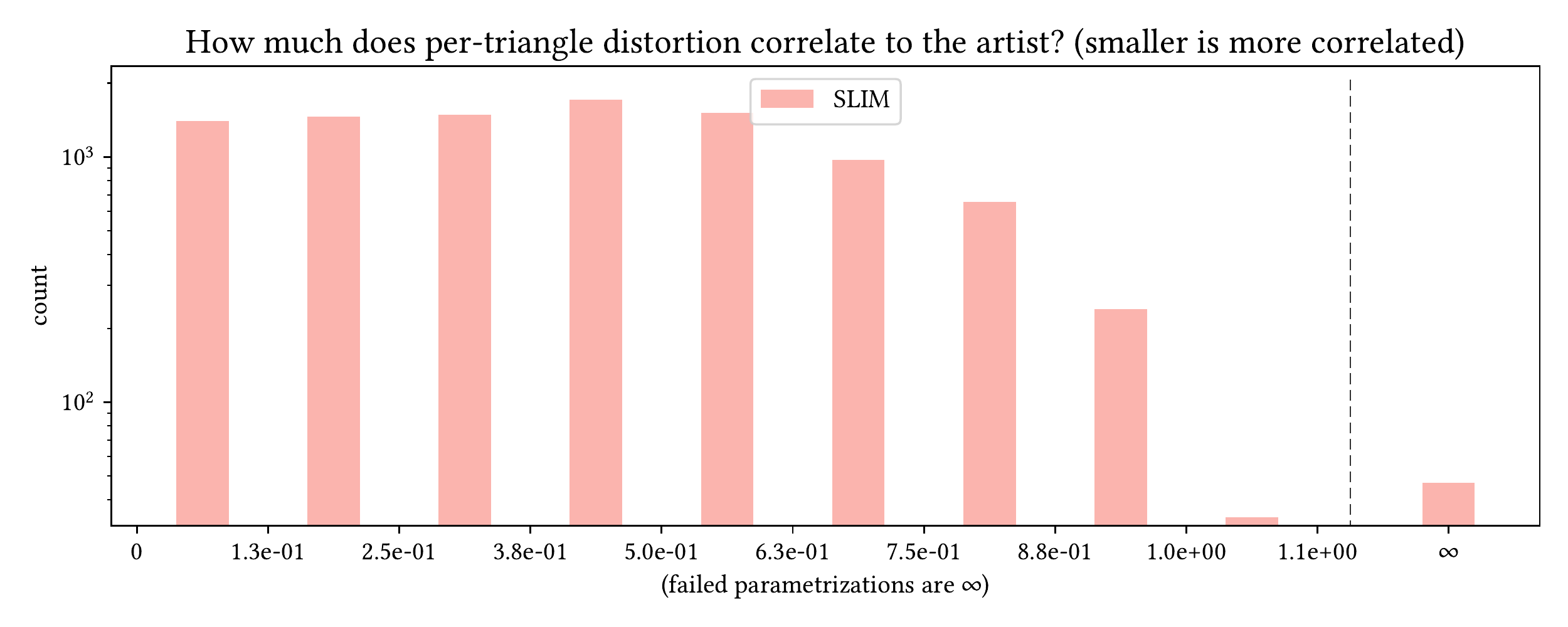}
    \includegraphics[width=\linewidth]{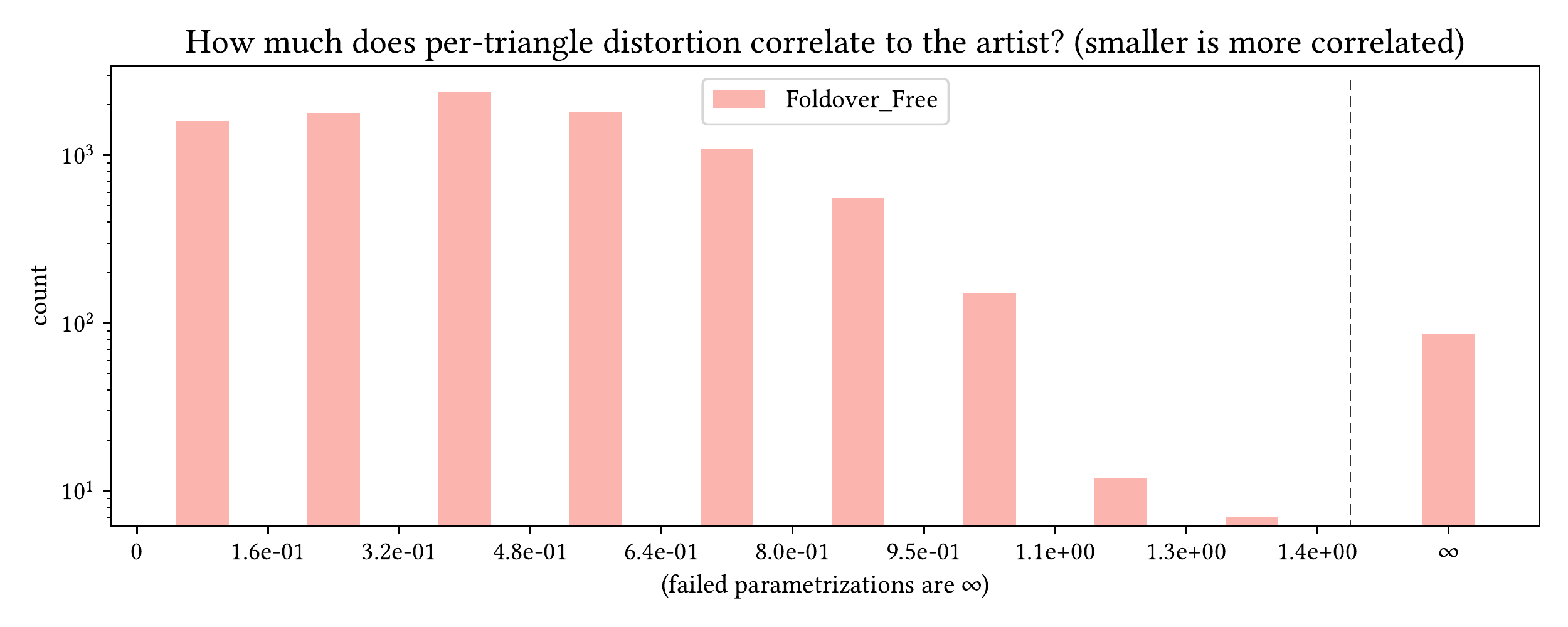}
    \caption{
    How much does the distortion of SLIM \cite{slim} and Foldover-Free
    \cite{foldover-free} correlate to the distortion of the artist on each mesh?
    Each mesh receives a correlation score from $0$ (perfect correlation, best) to
    $2$ (negative correlation, worst), and these scores are displayed in a
    histogram.
    The high values imply that these automatic methods distribute distortion
    very differently from artists.
    }
    \label{fig:correlation}
\end{figure}

Fig.~\ref{fig:correlation} shows that neither SLIM nor Foldover-Free produce UV
maps whose per-triangle distortion correlates with the artist's.
This is not surprising, since that is not the goal of these automatic
parameterization methods.
It is, however, interesting to highlight the large difference between UV maps
produced by artists and UV maps produced by automatic methods:
these two automatic methods do not seem to produce results that match artist
intent in per-triangle distortion.
Producing UV maps that match artist intent is a promising  research direction
for future data-driven parameterization methods, and we hope that our benchmark will
be a useful tool in this endeavor.

\section{Limitations}

Our benchmark does not capture all important aspects of parameterization methods.
We rely on provided parameterized meshes and as such do not have access to information about the runtime of the parameterization method on each mesh.
Users have to record and report this information themselves.

We cannot be sure which aspects of an artist's UV map are intentional.
Our comparison to the hand-generated parameterizations aims to reveal
what characteristics of a parameterization are important to the users of such a
parameterization.
However, just because a parameterization was produced by an artist does not mean
it would be the preferred parameterization for the artist if they had access
to other texturing methods.
Artists might also use some UV map generated by their modeling software, and
not try to improve it further if it is already good enough for their
applications.
Classifying the artist's intention is left to future work.

\section{Conclusion}

We provide a large dataset of real-world meshes with included artist-provided texture maps.
Our meshes have wildly varying topology and thus provide a challenging measure of parameterization methods' robustness to the complexity and quality of meshes in the wild.
We also provide a benchmark script that compiles standardized statistics for the UV maps produced by a parameterization method on these meshes.

Our benchmark presents many opportunities for future work.
There are other metrics worth exploring for inclusion in the benchmark,
such as a measure for UV area overlap,
which could provide a measure of global injectivity.

Another avenue of future work is in examining the specific applications of the meshes in our dataset. The meshes in our dataset come from a variety of sources, which is an advantage in testing for method robustness. These applications, however, require different properties for a parameterization method. By tagging the dataset with the intended use---such as 3D printing, game assets, or photorealistic rendering---we could
classify the meshes by application.

\section{Acknowledgements}
The MIT Geometric Data Processing group acknowledges the generous support of Army Research Office grants W911NF2010168 and W911NF2110293, of Air Force Office of Scientific Research award FA9550-19-1-031, of National Science Foundation grants IIS-1838071 and CHS-1955697, from the CSAIL Systems that Learn program, from the MIT--IBM Watson AI Laboratory, from the Toyota–CSAIL Joint Research Center, from a gift from Adobe Systems, from an MIT.nano Immersion Lab/NCSOFT Gaming Program seed grant, and from a Google Research Scholar award.
This work is supported by the Swiss National Science Foundation’s Early Postdoc.Mobility fellowship.
We thank Mazdak Abulnaga, Yu Wang, and Lingxiao Li for help with proofreading.

\bibliographystyle{eg-alpha-doi} 
\bibliography{main}

\end{document}